\documentclass{article}

\usepackage[preprint]{arxiv_2023}

\usepackage[utf8]{inputenc} % allow utf-8 input
\usepackage[T1]{fontenc}    % use 8-bit T1 fonts
\usepackage{hyperref}       % hyperlinks
\usepackage{url}            % simple URL typesetting
\usepackage{booktabs}       % professional-quality tables
\usepackage{amsfonts}       % blackboard math symbols
\usepackage{nicefrac}       % compact symbols for 1/2, etc.
\usepackage{microtype}      % microtypography
\usepackage{xcolor}         % colors
\usepackage{multirow}
\usepackage{graphicx}
\usepackage{makecell}
\usepackage{subfigure}
\usepackage{fontawesome}
\usepackage{caption} 

\makeatletter
\newcommand\figcaption{\def\@captype{figure}\caption}
\newcommand\tabcaption{\def\@captype{table}\caption}
\makeatother

\begin{document}
% \title{A Benchmark Dataset for Large-Scale Micro-Video Recommendation}
\title{A Content-Driven Micro-Video Recommendation Dataset at Scale}

% The \author macro works with any number of authors. There are two commands
% used to separate the names and addresses of multiple authors: \And and \AND.
%
% Using \And between authors leaves it to LaTeX to determine where to break the
% lines. Using \AND forces a line break at that point. So, if LaTeX puts 3 of 4
% authors names on the first line, and the last on the second line, try using
% \AND instead of \And before the third author name.

% \author{%
%   David S.~Hippocampus\thanks{Use footnote for providing further information
%     about author (webpage, alternative address)---\emph{not} for acknowledging
%     funding agencies.} \\
%   Department of Computer Science\\
%   Cranberry-Lemon University\\
%   Pittsburgh, PA 15213 \\
%   \texttt{hippo@cs.cranberry-lemon.edu} \\
%   % examples of more authors
%   % \And
%   % Coauthor \\
%   % Affiliation \\
%   % Address \\
%   % \texttt{email} \\
%   % \AND
%   % Coauthor \\
%   % Affiliation \\
%   % Address \\
%   % \texttt{email} \\
%   % \And
%   % Coauthor \\
%   % Affiliation \\
%   % Address \\
%   % \texttt{email} \\
%   % \And
%   % Coauthor \\
%   % Affiliation \\
%   % Address \\
%   % \texttt{email} \\
% }

\author{
Yongxin Ni\textsuperscript{\rm 1},~Yu Cheng\textsuperscript{\rm 1},~Xiangyan Liu\textsuperscript{\rm 1},~Junchen Fu\textsuperscript{\rm 1},~Youhua Li\textsuperscript{\rm 1},\\~\textbf{Xiangnan He}\textsuperscript{\rm 2},~\textbf{Yongfeng Zhang}\textsuperscript{\rm 3},~\textbf{Fajie Yuan}\textsuperscript{\rm 1}\thanks{Corresponding author. Author contributions: Fajie designed and supervised this research; Yongxin performed the research including key experiments; Chengyu, Junchen, Youhua, Xiangyan assisted a few important experiments; Xiangnan and Yongfeng provided guidance, participated in discussions, and proofread the paper; Fajie and Yongxin led the paper writing.}\\
\textsuperscript{\rm 1}Westlake University 
\\
\texttt{\{niyongxin,chengyu,liuxiangyan,fujunchen,liyouhua,yuanfajie\}@westlake.edu.cn}
\\
\textsuperscript{\rm 2}University of Science and Technology of China
\\
\texttt{\{xiangnanhe\}@@gmail.com}
\\
\textsuperscript{\rm 3}Rutgers University 
\\
\texttt{\{yongfeng.zhang\}@rutgers.edu}
}

\maketitle

\begin{abstract}
Micro-videos have recently gained immense popularity, sparking critical research in micro-video  recommendation with significant implications for the entertainment, advertising, and e-commerce industries. However, the lack of large-scale public micro-video  datasets poses a major challenge for developing effective recommender systems. To address this challenge, we introduce a very large micro-video recommendation dataset, named ``\emph{MicroLens}'', consisting of one billion  user-item interaction behaviors, 34 million  users, and one million  micro-videos. This dataset also contains various raw modality information about videos, including titles, cover images, audio, and full-length videos. MicroLens serves as a benchmark for content-driven micro-video recommendation, enabling researchers to utilize various modalities of video information for recommendation, rather than relying solely on item IDs or off-the-shelf video features extracted from a pre-trained network.  
Our benchmarking of multiple recommender models and video encoders on MicroLens has yielded valuable  insights into the performance of micro-video recommendation. We believe that this dataset will not only benefit the recommender system community but also promote the development of the video understanding field. Our datasets and code are available at  \url{https://github.com/westlake-repl/MicroLens}.
% In this paper, we benchmark multiple recommender models and video encoders on this dataset, revealing various insights.
% while examining different factors that impact recommendation performance. 

% Our findings have important implications for designing effective recommendation systems that take into account the rich features of micro-videos and the complex interplay between video understanding and recommendation. We anticipate that this dataset will facilitate further research in this area and help advance the state-of-the-art in video recommendation.
\end{abstract}
% Micro-video recommendation is a critical problem with numerous applications and a vast user base. However, the scarcity of large-scale datasets containing a diverse range of raw modality features has resulted in inadequate research on user profiling and micro-video recommendation. To this end, we present MVRec, a large-scale modality-based benchmark dataset for \underline{m}icro-\underline{v}ideo \underline{rec}ommendation. It consists of 30 milMicroLens users, 1 milMicroLens videos, and 1 bilMicroLens interactions, along with raw multi-modal features such as title, cover, audio, and video for all items. Apart from validating our MVRec using traditional recommendation baselines, we conduct a comprehensive evaluation by benchmarking 15 pre-trained video understanding models on the proposed dataset. Our findings reveal that an improved vision model does not inherently ensure enhanced representation performance when predicting human behavior. Furthermore, we assess the transferability of our dataset by integrating additional video datasets from various platforms. Our results indicate that the proposed datasets can serve as valuable resources for researchers to investigate user representation and modality-based recommendation. We make MVRec and all associated downstream datasets publicly available, complete with their corresponding raw multi-modal features.
% \end{abstract}

\section{Introduction}
Micro-videos, also known as short-form videos, have become increasingly popular in recent years. These videos typically range in length from a few seconds to several minutes and exist on various platforms, including social media, video-sharing websites, and mobile apps. Due to the brief yet captivating content, micro-videos have captured the attention of audiences worldwide, making them a powerful means of communication and entertainment. The surge in popularity of micro-videos has fueled critical research in micro-video recommender  systems~\cite{zheng2022dvr,yu2023improving,gong2022real,yuan2022tenrec,gao2022kuairec,liu2021concept,han2016dancelets,liu2019user,jiang2020aspect,lei2021semi}.

However, the absence  of large-scale public micro-video datasets containing diverse and high-quality video content, along with user behavior information, presents a significant challenge in developing reliable  recommender systems (RS).
Existing video recommendation datasets, such as MovieLens~\cite{harper2015movielens}, mainly focus on longer movie-type videos  and do not cover the wide range of content found in micro-videos, including but not limited to categories such as food, animals, sports, travel, education, fashion, and music.
Additionally, other datasets such as Tenrec~\cite{yuan2022tenrec} and KuaiRec~\cite{gao2022kuairec} only contain video ID data or pre-extracted vision features from the video thumbnails, making it difficult to develop recommender models that can learn video representation directly from the raw video content data. Thus, there is an urgent need for large-scale micro-video recommendation datasets offering diverse raw content to facilitate the development of more accurate and effective recommender algorithms.

% Although there are many video datasets available, most of them focus on longer videos and do not contain the diverse range of content found in micro-videos. For example, the MovieLens~\cite{harper2015movielens} dataset only contains videos related to movies and does not include other categories such as food, animals, sports, or shows.
% Other datasets, such as TenRec~\cite{yuan2022tenrec} and KuaiRec~\cite{gao2022kuairec}, only contain video ID data or pre-extracted vision features from the video thumbnails. These datasets lack the rich and diverse set of video content found in micro-videos, making it challenging to develop effective recommendation systems that can account for the various modalities of video information, such as audio, metadata, and other visual features. Therefore, the need for a large-scale micro-video dataset that contains a diverse range of high-quality video content has become increasingly pressing for researchers in this field.

 % Therefore, the need for a large-scale micro-video dataset that contains a rich and diverse set of video content has become increasingly urgent. Such a dataset would enable researchers to develop more robust and accurate recommendation algorithms that utilize various modalities of video information, rather than relying solely on item IDs or pre-extracted feature representations.

To address this challenge, we introduce a large-scale micro-video recommendation dataset, named ``\emph{MicroLens}'', consisting of one billion  user-item interaction behaviors, 34 million users, and one million micro-videos.  Each micro-video is accompanied by original modalities, such as title, cover image, audio, and video information, providing a rich and diverse set of features for recommender models. 
% benchmark for personalized micro-video recommendation, providing researchers with a valuable resource for developing and evaluating  video  content-driven recommender models. 
% Our dataset enables researchers to utilize various modalities of video information for recommendation, rather than relying solely on item IDs or pre-extracted feature representations. 
Then, we perform benchmarking of various recommender baselines and cutting-edge video encoders on this dataset, providing valuable insights into the recommendation accuracy.
We believe MicroLens can serve as a valuable resource for developing and evaluating    content-driven video recommender models. 
 % Moreover, if we treat video recommendation as a downstream task for video understanding, MicroLens can be also useful to advance the research on   video understanding. 
 To summarize, our contribution in this paper is three-fold: 
\begin{itemize}
\item We introduce the  largest and most diverse 
micro-video recommendation dataset, which provides access to raw video data. MicroLens encompasses all important modalities, including image, audio, text, and full-length video,
making it an ideal resource for researchers working in various areas related to multimodal recommendation.
 % MicroLens can serve as a valuable resource for researchers working in diverse fields related to video, image, text, and multimodal recommendation, enabling them to explore various modalities of video information for recommendation purposes.
\item We provide a comprehensive benchmark for over 10 recommender models and video encoders.  Additionally, we introduce new types of baselines that use end-to-end (E2E) training to optimize both recommender models and video encoders. Although computationally expensive, these E2E models achieve superior performance that remains unknown in literature.
% \textcolor{blue}{since you present a new dataset, do you have any new findings? I believe high-cost is not the core contribution. }
\item Through empirical study, we present several crucial insights and explore the potential relationship between video understanding and recommender systems. Our findings indicate that a significant gap exists between current video understanding technologies and video recommendation, emphasizing the need for specialized research on video understanding technologies for video recommendation tasks.
% \textcolor{blue}{seems to be vague.}
\end{itemize}

\section{MicroLens}

\subsection{Dataset Construction}

\textbf{Seed Video Collection.} 
The data for MicroLens is sourced from an online micro-video platform with a focus on social entertainment. The recommendation scenario is described in Appendix Figure~\ref{recommendation-scenario-figure}.
% \footnote{The platform name has been anonymous here
% suggested by the company.}
% The data for MicroLens is entirely sourced from Kuaishou\footnote{https://www.kuaishou.com/?isHome=1}, which is a well-known micro-video sharing platform in China with a primary focus on entertainment content. 
The data collection process spanned almost a year, from June 2022 to June 2023. To begin with, we collected a large number of seed videos from the homepage. To ensure the diversity of videos, we frequently refreshed the homepage, allowing us to obtain a new set of videos every time. To ensure the quality of the collected videos, we filtered out unpopular  content by only including videos with more than 10,000 likes in this stage. It is important to note that the platform does not  directly provide user-video like and click interactions due to privacy protection. Instead, it provides  user-video comment behaviors, which are publicly available and  can serve as an implicit indicator of strong user preference towards a video. On average, there is approximately  one  comment for  every 100 likes. That is, we mainly collected videos with  positive interactions greater than 100  to ensure a reasonable level of engagement.\footnote{If the items are too cold in the platform, it is almost impossible to find enough overlapping users.}  In total, we collect 400,000 micro-videos, including their video title, cover image, audio and raw video information.

\textbf{Dataset Expansion.} 
In this stage, we accessed the webpages of the videos collected in the previous stage. Each video page contains numerous links to external related videos, from which we randomly selected 10 video links. Note that (1) the related video links on each video page change with each visit; (2) the related videos have very diverse themes and are not necessarily of the same category as the main video. We collected approximately 5 million videos in this stage and retained the same metadata as in the previous stages.

% To prevent collecting videos that may be platform-directed based on previous click behavior, we used a new cookie for each video visit and refrained from clicking on any videos. We collected approximately 4 million videos in this stage and retained the same metadata as in the previous stages.

% Following the acquisition of a substantial number of random videos, the video set was further expanded by leveraging the "Related Videos" section found on each video's detail page, which presented an additional 20 videos bearing stylistic similarities to the current video. To maintain the balance between high relevance and sufficient expansion, we adopt a different selection strategy: we collect the top ten videos from this section without imposing a like-number threshold. To ensure the integrity of our data and avoid potential platform-induced bias, we utilized a new coolie every time we accessed a video, ensuring no historical click behavior was recorded prior to each video visit, and consciously refraining from interacting with any videos within the "Related Videos" section. Throughout this expansion phase, we collected eight hundred thousand videos totally, and preserved the same type of metadata as the first phase.

\textbf{Data Filtering.}
After collecting the videos, we conducted a data filtering process to remove a large number of duplicates and filter the data based on different modalities. For the text modality, we required that the length of the video titles, after removing meaningless characters, should not be less than 3. For the image modality, we used color uniformity checks and removed images with single-color areas greater than 75\%. For the video modality, we set a threshold for file size and removed any videos with a file size of less than 100KB.

Overall, these filtering criteria helped to improve the quality of the collected data and ensured that only relevant and high-quality videos were included in the final dataset.
% Upon the completion of the two-stage collection process, we applied a filtering mechanism to the obtained videos to remove duplicates, thereby finalizing a comprehensive and unique set of videos. Also, data with incomplete modalities were systematically filtered out to maintain the integrity of the dataset. In addition to this, each modality was subjected to distinct filtering methods to ensure the quality of multimodal data. For the text modality, a process was implemented to remove characters devoid of semantic meaning, with a stipulation that the resultant length should not be less than three. In the case of the image modality, images were subjected to a color uniformity check, and images with a single color region exceeding $75\%$ were excluded. For the video modality, a size threshold was set, and videos smaller than 100k were filtered out. These measures were meticulously implemented to ensure the quality and reliability of our multimodal MicroLens dataset.

\textbf{Interaction Collection.} 
In this stage, we collected user-video interaction behaviors, primarily through the collection of comment data. We chose to collect comment data as a form of positive feedback from users for two primary reasons: (1) all user comment data on the platform is public, which eliminates potential privacy concerns that may arise with click and like data;
% \footnote{User click and like behaviors can only be obtained from the logs of recommender systems within the company. Since they are non-public data, the publishing process may involve legal issues such as privacy and data security.} 
and (2) unlike e-commerce scenarios where negative comments often indicate user dissatisfaction with the product, comments on short videos are typically about the people and events portrayed in the video, and both positive and negative comments can serve as indicators of user preferences towards the video. In fact, these preferences may be even stronger than those inferred from click behaviors.

To collect comments, we accessed the webpage of each video and collected up to 5000 comments per video. This limitation was due to the fact that collecting more comments through pagination would require more time. In addition, we removed multiple comments from the same user to ensure data quality. Apart from comment data, we also recorded user IDs and comment timestamps. Although user and video IDs are public, we still anonymize them to avoid any privacy concerns.
% However, for further privacy protection, we anonymized both the user IDs and video IDs.
% we recorded up to 6500 interactions and ignored multiple interactions from
% the same user.
% User feedback was collected by accessing the comment page of each video and obtaining the top 5000 published comments under each video, including the user ID of the commenter  and the corresponding timestamp. The collection of user-video interactions based on comments was driven by two primary considerations. Firstly, given the public visibility of comments, the collection does not infringe upon user privacy. To further ensure privacy, the specific content of comments was not collected. Instead, only the user IDs of the commenter were procured, and a further privacy protection was conducted by anonymizing both the user and video IDs. Secondly, comments provide a deeper insight into user intentions compared to the more prevalent click behavior. This is predicated on the understanding that users are compelled to open the video and leave a comment only when they are sufficiently moved to express their views. The MicroLens dataset was established based on the collected multimodal data and user-video commenting interactions, comprehensively offering a panoramic view of user-video interactions, thereby providing valuable insights into user behavior and preferences.

\begin{figure*}[t]
    \centering
    \includegraphics[width=0.98\textwidth]{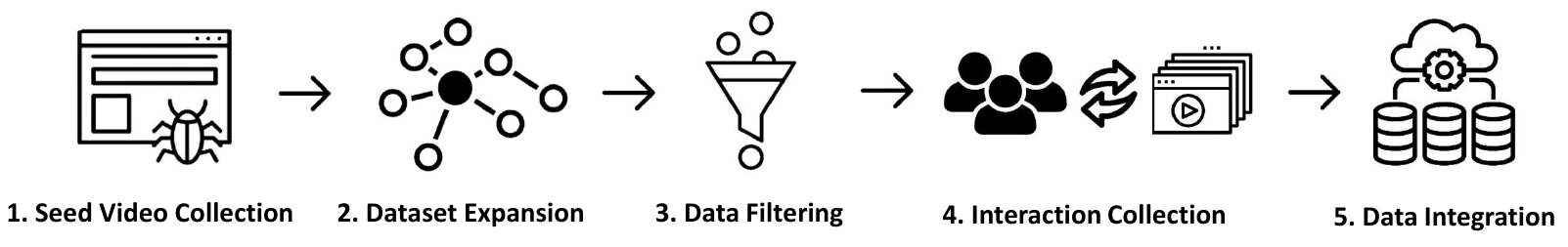}
    \caption{Dataset construction pipeline.}
    \label{dataset-construction}
\end{figure*}

\begin{figure}[t]
    \centering
    \includegraphics[width=\textwidth]{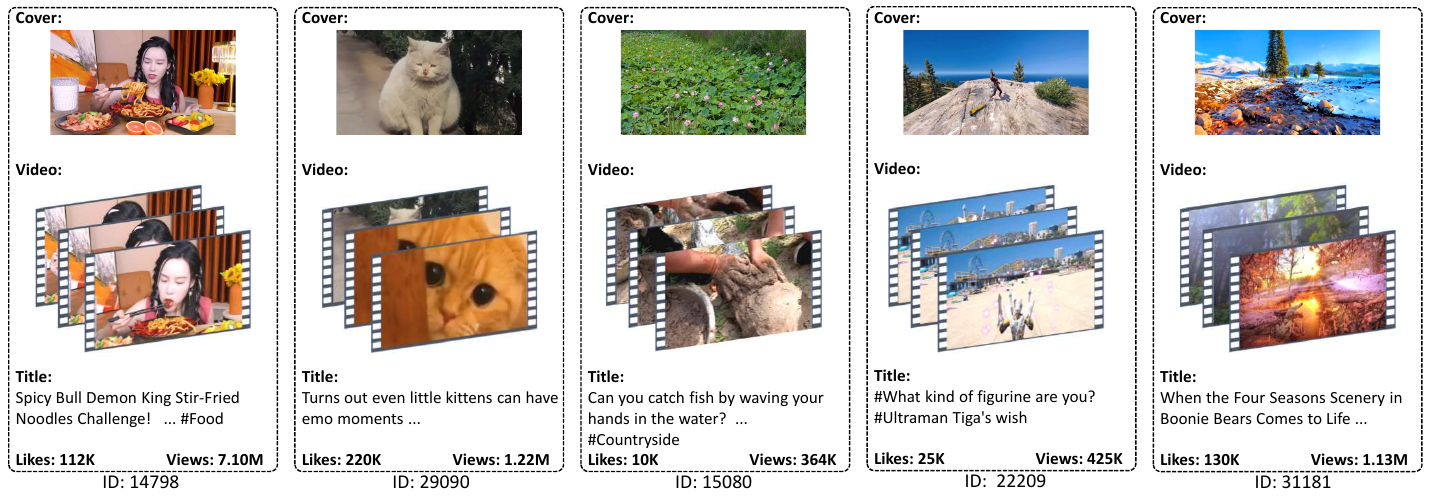}
    \caption{Item examples in MicroLens.}
    \label{samples-display}
\end{figure}

\textbf{Data Integration.} 
Due to the large volume of collected data, we employed a specialized data integration process. Our approach involved using a distributed large-scale download system,  consisting
of collection nodes, download nodes, and a data integration node. The technical details are 
in Appendix Section~\ref{technical-details}. We provide the dataset construction process in Figure \ref{dataset-construction}, and samples in Figure~\ref{samples-display}.

\subsection{Privacy and Copyrights}
MicroLens only includes public user behaviors for privacy protection. Both user  and item IDs have been anonymized. To avoid copyright issues, we provide video URLs and a special tool to permanently access and download related videos. This is a common practice in prior literature ~\cite{nielsen2022mumin,zeng2022tencent} when publishing multimedia datasets, e.g. YouTube8M\footnote{Note that YouTube8M does not include user interaction data and therefore is not a recommendation dataset.}~\cite{abu2016youtube}. We will also provide the original dataset with reference to the ImageNet license, see \url{https://www.image-net.org/download.php}.

% \begin{figure*}[t]
% \centering
% \begin{minipage}[b]{0.32\textwidth}
%     \includegraphics[width=\textwidth]{figure/item-pop-density.png}
%     \vspace{-0.04in}
%     \caption{Item popularity.}
%     \label{fig:popularity-figure}
% \end{minipage}
% \hfill
% \begin{minipage}[b]{0.32\textwidth}
%     \includegraphics[width=\textwidth]{figure/user-frequency.png}
%     \caption{User activity.}
%     % \vspace{0.13in}
%     \label{fig:activity-figure}
% \end{minipage}
% \hfill
% \begin{minipage}[b]{0.32\textwidth}
%     \includegraphics[width=\textwidth]{figure/video-duration.png}
%     \caption{Video duration.}
%     % \vspace{0.13in}
%     \label{fig:duration-figure}
% \end{minipage}
% \end{figure*}

% \begin{figure*}[h]
%     \begin{minipage}{0.35\columnwidth}
%       \includegraphics[width=\linewidth]{figure/item-pop-density.png}
%        (a) cumulative density v.s. video rank
%     \end{minipage}
%     \begin{minipage}{0.32\columnwidth}
%       \includegraphics[width=\linewidth]{figure/user-frequency.png}
%        (b) frequency v.s. user session length
%     \end{minipage}
%     \begin{minipage}{0.32\columnwidth}
%       \includegraphics[width=\linewidth]{figure/video-duration.png}
%        (c) frequency v.s. video duration
%     \end{minipage}
%     \caption{}
%     \label{fig:stat}
% \end{figure*}

\subsection{Dataset Analysis}
% Table \ref{sts-table}  provides relevant statistics of our dataset, highlighting the key characteristics of MicroLens. The dataset comprises xx users, xx items, and xx comment interactions. 
% On average, each user interacted with xx items (denoted by \textit{U\_pop}), and each item interacted with xx users (denoted by \textit{I\_pop}).
% Each user participated in an average of xx interactions (\textit{U\_pop}), with a mean time span of xx months between their first and last interactions (\textit{U\_span}). Items in the dataset exhibit average popularity of xx interactions by users (\textit{I\_pop}), and the mean span a video is commented on is 2 months (\textit{I\_span}). Each item features a title in both English and Chinese, with an average length of xx and xx characters, respectively (\textit{Length}). \textit{Tags} that imply the videos' subdivided categories are included in these titles, and our MicroLens covers total xx unique tags. These micro-videos' average length is xx seconds (\textit{Duration}), indicating their short-duration and content-concentrated characteristics.

% This suggests that a short clip is typically sufficient for carrying out video understanding tasks on MicroLens.

% As the original MicroLens dataset is too large for most academic research, we have created two subsets of MicroLens by randomly selecting 100,000 users and 1 million users. These subsets are named MicroLens-100K  and MicroLens-1M respectively. We consider MicroLens-100K as the default dataset for academic research and provide additional results on MicroLens-1M in the Appendix.

As the original MicroLens dataset is too large for most academic research, we have created two subsets of MicroLens by randomly selecting 100,000 users and 1 million users, named MicroLens-100K and MicroLens-1M, respectively. We consider MicroLens-100K as the default dataset  to evaluate recommender models and provide some key results on MicroLens-1M in the Appendix. 
% \textcolor{blue}{Because it is similar to MovieLens, do you have some comparison on the basic  statistics? MovieLens did not provide audio and video modality, we have fixed the comparison table.}
\begin{figure}
  \centering
  \subfigure[Item popularity.]{
    \begin{minipage}[b]{0.3\textwidth}
    \includegraphics[width=\linewidth]{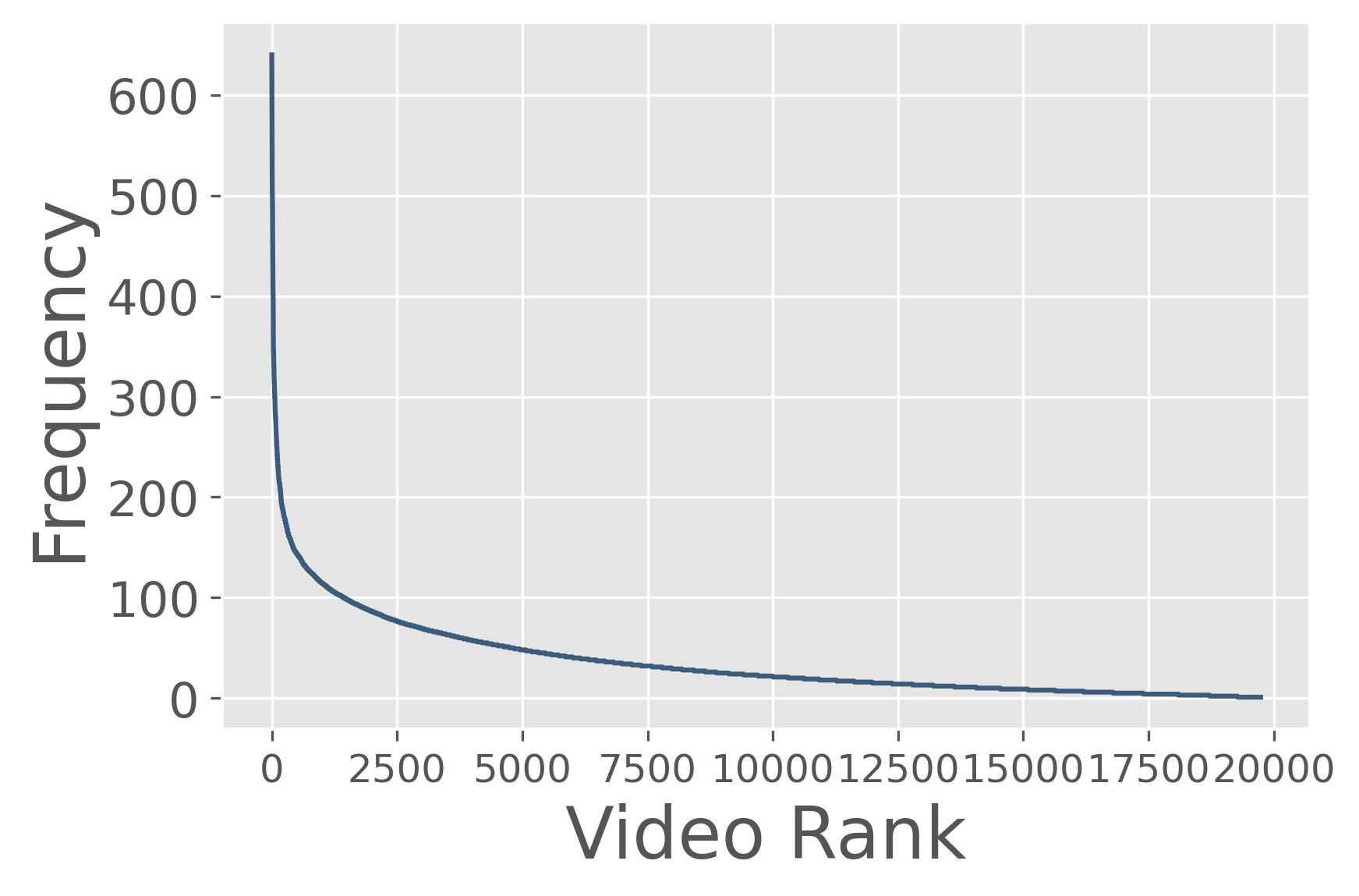}
    \vspace{-0.04in}
    
    \end{minipage}
    \label{fig:popularity-figure}
  }
  \hfill
  \subfigure[User session length.]{
    \begin{minipage}[b]{0.3\textwidth}
    \includegraphics[width=\linewidth]{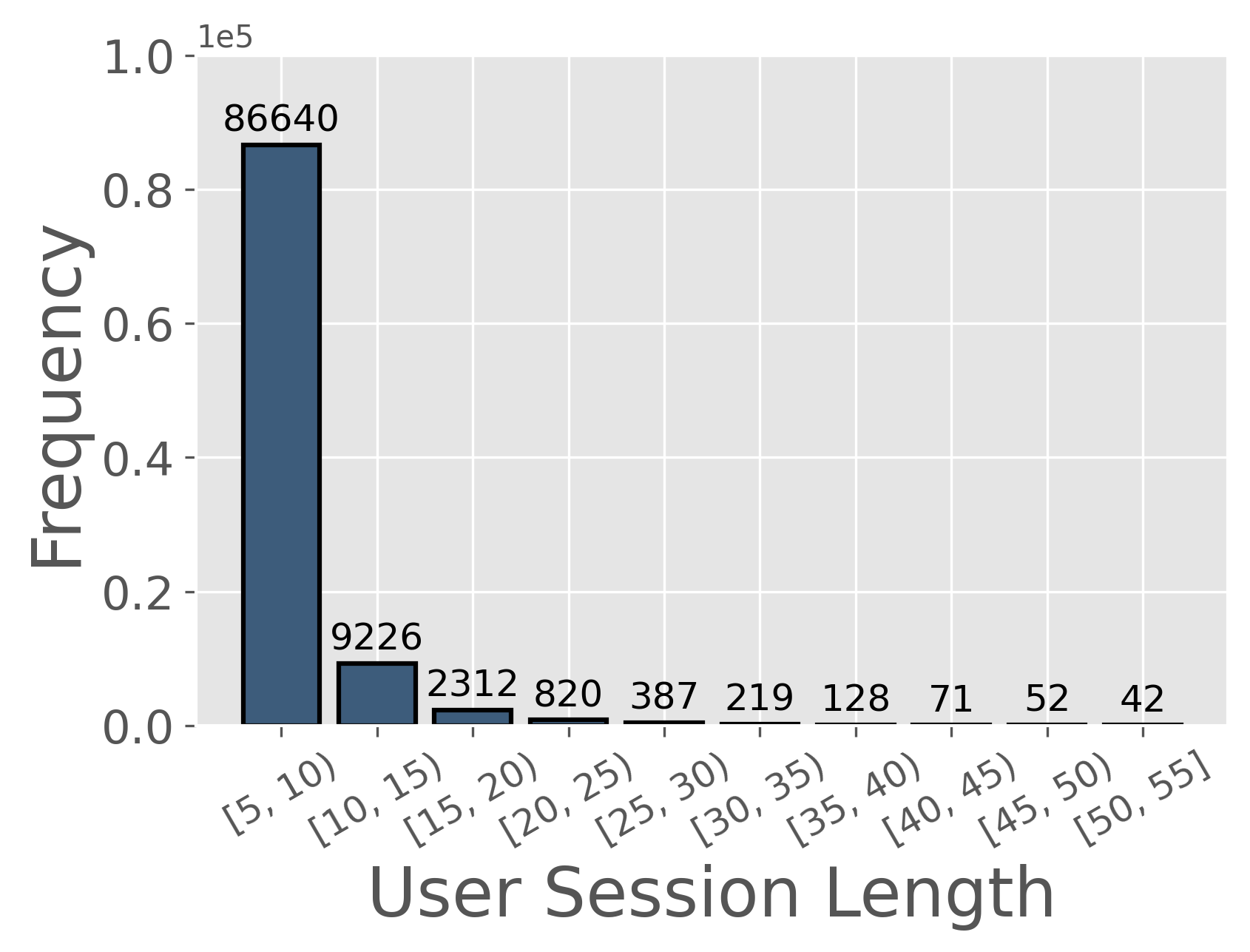}
    
    \end{minipage}
    \label{fig:activity-figure}
  }
  \hfill
  \subfigure[Video duration  (in seconds)]{
    \begin{minipage}[b]{0.3\textwidth}
    \includegraphics[width=\linewidth]{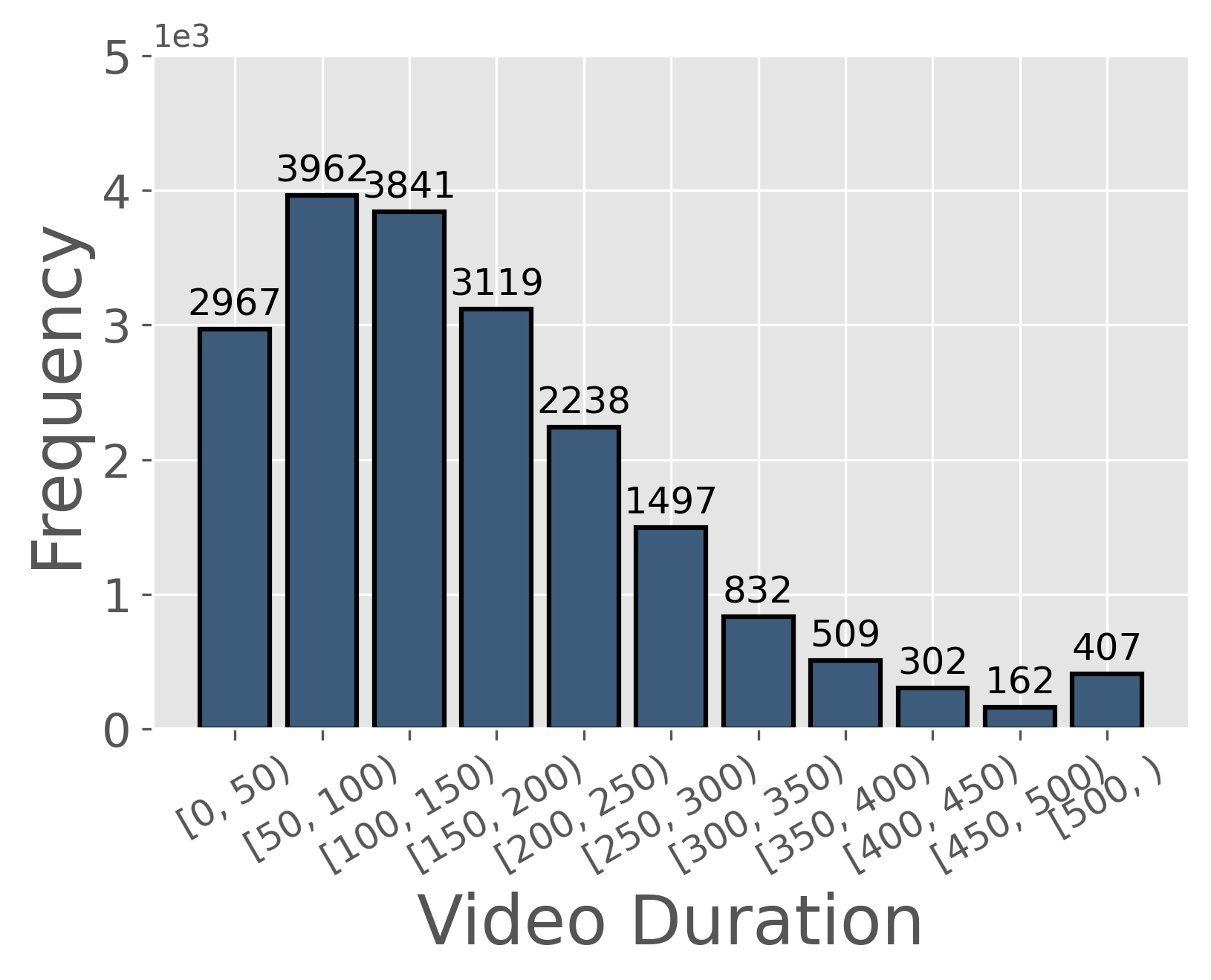}
    
    \end{minipage}
    \label{fig:duration-figure}
  }

  \caption{Statistics of MicroLens-100K.}
  \label{fig:distribution}
\end{figure}

Figure \ref{fig:distribution} illustrate some statistics of MicroLens-100K.
% \footnote{We put the illustration figures of MicroLens-1M and MicroLens in Appendix Section~\ref{larger-datasets}.} 
(a) shows that item popularity aligns with the long-tail distribution which is commonly observed in most
recommender systems. 
% The red dotted line in Figure (a) represents the median of video rank, it intersects with the cumulative density line at 0.17, indicating that the top 50\% of popular items contributed 83\% of the total interaction volume. 
(b) indicates that users with interaction sequence lengths between 5 and 15 constitute the majority group.
% , and approximately 5\% of the users possess a sequence length that exceeds 15.
% rendering MicroLens a valuable resource for studying users' cold-start problem.
(c) depicts the distribution of video duration, with the majority of micro-videos less than 400 seconds in length. 

We present the detailed statistical information of MicroLens-100K, MicroLens-1M, and the original MicroLens in Table~\ref{sts-table}. MicroLens-100K comprises 100 thousand users, 19,738 items, and 719,405 interactions, with the sparsity of 99.96\%. MicroLens-1M includes 1 million users, 91,402 items, and 9,095,620 interactions, with the sparsity of 99.99\%. The original MicroLens dataset consists of 34,492,051 users, 1,142,528 items, and 1,006,528,709 interactions.
% The average lengths of their Chinese and English video titles are 38/xx, 41/xx, and 47/xx characters, respectively. 
The three datasets, in ascending order, contain 15,580, 28,383, and 258,367 tags, respectively, with each tag representing a fine-grained category to which the videos belong. 
% Their average video durations are 161, 162, and xx seconds respectively, which is much shorter than the videos in the long movie dataset.
In addition to the raw multimodal information, we have also included additional features such as the number of views and likes per video, user gender information, and comment content.
% indicating the short-duration and content-concentrated characteristics of micro-videos.
% Note that videos of all datasets encompass a variety of original modalities, including video, audio, image, and text (abbreviated as VAIT in the table).

\begin{table*}[t]
    \caption{Data statistics of MicroLens.
    VAIT represents the video, audio, image  and text data.
}
    \label{sts-table}
    \centering
    % \vskip 0.12in
    \scalebox{0.9}{
    \begin{tabular}{lccccccc}
        \toprule
        Dataset & \#User & \#Item & \#Interaction & Sparsity & \#Tags & Duration & VAIT  \\
        \midrule
        MicroLens-100K & 100,000 & 19,738 & 719,405 & 99.96\% & 15,580 & 161s & \textcolor{green}{\faCheck} \\
        MicroLens-1M & 1,000,000 & 91,402 & 9,095,620 & 99.99\% & 28,383 & 162s & \textcolor{green}{\faCheck} \\
        MicroLens & 34,492,051 & 1,142,528 & 1,006,528,709 & 99.997\% & 258,367 & 138s & \textcolor{green}{\faCheck} \\
        \bottomrule
    \end{tabular}
    }
\end{table*}

\subsection{Comparison to Existing Datasets}
Over the past two decades, the field of RS  has accumulated a large number of benchmark datasets. The most representative of these, MovieLens~\cite{harper2015movielens}, has been extensively utilized for various recommendation tasks, particularly the rating prediction and top-N item recommendation tasks. Additionally, both academia and industry have released high-quality datasets, including Alibaba's various CTR prediction datasets~\cite{zhou2018deep, zhu2018learning}, Tencent's Tenrec dataset~\cite{yuan2022tenrec}, and Kuaishou's KuaiRec and KuaiRand datasets~\cite{gao2022kuairec, gao2022kuairand}.  However, the majority of public RS datasets only offer user IDs, item IDs, and click behaviors, with relatively few public datasets providing multimodal information about the items. While KuaiRec, Flickr~\cite{wu2019hierarchical}, and Behance~\cite{he2016vista} offer multimodal features, it is noteworthy that the image features are pre-extracted from vision encoders (e.g. ResNet~\cite{he2016deep}) without raw pixel features. Recently, Microsoft released the MIND dataset~\cite{wu2020mind}, which is the largest news recommendation dataset to date. Amazon~\cite{he2016ups} and POG~\cite{chen2019pog} provided a large product purchase dataset that includes raw images of products. In addition, H\&M\footnote{https://www.kaggle.com/competitions/h-and-m-personalized-fashion-recommendations}, Yelp\footnote{https://www.yelp.com/dataset}, and GEST~\cite{yan2022personalized} (Google restaurant) also released several  image datasets for business recommendation purposes. A list of related multimodal datasets in the recommender system community are shown in Appendix Section \ref{dataset-comparision}.

% Some well-known multimodal datasets include Microsoft's MIND dataset for text and multimodal data, Amazon's product purchase dataset, H\&M's clothing purchase dataset, and Yelp and Google Restaurant's POI dataset. 

However, to our best knowledge, there is currently no micro-video recommendation dataset that provides original video content. Reasoner~\cite{chen2023reasoner} and MovieLens are two relevant datasets to MicroLens. However, the size of the Reasoner dataset is significantly smaller and only offers five frames of images for each video, whereas our MicroLens includes about 10,000 times more users and user-video behaviors.  Although MovieLens provides the URLs of the movie trailer, it has a limited range of video categories, as it only includes videos of movie categories. Additionally, MovieLens was collected from a simulated user-movie rating website and does not accurately represent actual user watching behavior. For instance, in MovieLens, many users can rate over 10 movies in a matter of seconds, which does not reflect actual movie watching behaviors.  As a result, despite its frequent use, it may not be ideal for certain types of research, e.g., sequential recommendation. Please refer to \cite{woolridge2021sequence} for in-depth analysis of MovieLens.

\section{Experiments Setup}
Despite MicroLens's potential for multiple research areas, our primary focus here is on the micro-video recommendation task. As mentioned, we chose MicroLens-100K as our default dataset for basic evaluation, while additional results using MicroLens-1M are reported in Appendix.
% \subsection{Experimental Setup}

% \textbf{Baselines and Evaluation.}
\subsection{Baselines and Evaluation}

% Due to memory limitations during training, we constructed interaction sequences for each user by selecting their ten most recent items. 

% As mentioned, we choose MicroLens-100K as the default dataset for conducting experiments. Since there are memory limitations during the training, for all sequential methods, we construct interaction sequences for each user by selecting their most recent 10 items. Following previous works \cite{he2017neural}, we adopt the leave-one-out strategy to split the datasets: the last item in the interaction sequence is used for evaluation, and the item before the last is used as validation while the rest are for training. 

According to prior literature, video recommender
models can be broadly categorized into two groups: those relying on pure item IDs (i.e., video content-agnostic)  and those that incorporate  pre-extracted video or visual features (from a frozen video encoder) along with item IDs. Among them, models based on pure item IDs (called IDRec) can typically be further divided into classical collaborative filtering (CF) models, such as   DSSM~\cite{huang2013learning}, LightGCN~\cite{he2020lightgcn}, DeepFM~\cite{guo2017deepfm} and NFM~\cite{he2017neuralfm}, and sequential recommendation (SR) models, such as GRU4Rec~\cite{hidasi2015session}, NextItNet~\cite{yuan2019simple}, and SASRec~\cite{kang2018self}. Regarding the latter models that utilize pre-extracted video features as side information, 
% such as YouTube~\cite{covington2016deep} and VBPR~\cite{he2016vbpr}\footnote{The idea of VBPR can be directly extended to video recommendation by substituting image features with video features. VBPR is equivalent to DSSM with video features when there are no hidden layers.},
IDs  continue to be the main features, except in very cold or new item recommendation scenarios. We simply call this approach  VIDRec.\footnote{In fact, research on VIDRec is relatively scarce  compared to text and image-based recommendations. Even the most widely recognized model, the YouTube model, primarily relies on video ID and other categorical features, without explicitly leveraging the original video content features.}  VIDRec can share a similar network architecture with IDRec, but with additional video features incorporated into the ID embeddings.

Beyond the above traditional baseline models, we also introduce a new family of recommender models called VideoRec. This model simply replaces the item ID  in IDRec with a learnable video encoder. Unlike VIDRec, VideoRec uses end-to-end (E2E) training to optimize both the recommender model and the video encoder simultaneously. With the exception of the item representation module (ID embedding vs. video encoder), all other components of VideoRec and VIDRec are identical.
% that is pre-trained on the video understanding task beforehand.
% Unlike VIDRec, VideoRec adopts end-to-end (E2E) training to simultaneously optimize both the recommender model and the video encoder. Other than the item representation module (ID embedding vs. video encoder), all other components of VideoRec and IDRec are kept exactly the same.
Although VideoRec  achieves the highest recommendation accuracy, it has not been studied in literature due to its high training costs. 
% In addition, VideoRec may offer other advantages over IDRec, such as in cross-domain recommendation, by representing items based on raw video contents instead of non-sharable ID features. 
% This approach may have the potential to lead to the development of a universal, one-for-all recommender model in the future.

In terms of training details, we exploit the in-batch softmax loss function ~\cite{yi2019sampling} widely adopted in both academic literature and industrial systems.
For evaluation, we utilized the leave-one-out strategy to split the datasets where the last item in the interaction sequence was used for evaluation, the item before the last was used for validation, and the remaining items were used for training. As nearly 95\% of user behaviors involve less than 13 comments, we limited the maximum user sequence length to  the most recent 13  for sequential  models.
We employ two popular rank metrics~\cite{yuan2019simple,kang2018self}, i.e., hit ratio  (HR@N) and normalized discounted cumulative gain (NDCG@N). Here, N was set to 10 and 20. 

\subsection{Hyper-parameter Tuning}
% \textbf{Hyper-parameter Tuning.} 
As IDRec is the most efficient baseline, we conducted a hyper-parameter search for IDRec as the first step. Specifically, we first extensively search two key hyper-parameters: the learning rate $\eta$ from a set of values $\{1e-5, 5e-5, 1e-4, 5e-4, 1e-3\}$ and the embedding size from a set of values $\{64, 128, 256, 512, 1024, 2048, 4096\}$. Batch sizes $b$ were also empirically tuned for individual models from a set of values $\{64, 128, 256, 512, 1024, 2048\}$. %The weight decay and dropout were set to 0.1 after tuning.  
With regards to VIDRec and VideoRec, we first applied the same set of hyper-parameters obtained from IDRec and then performed some basic searches around these optimal values. It is
worth mentioning that extensively tuning VideoRec is not feasible in practice, as it requires at least 10-50 times more compute and training time than IDRec.\footnote{A recent study ~\cite{yang2022gram}  proposed a highly promising solution to improve training efficiency, which appears to be feasible for VideoRec.} Due to the very high computational cost involved, we only optimized the top few layers for the video encoder network in VideoRec. Along with other hyper-parameters, such as the layer numbers of NextItNet, GRU4Rec, and SASRec, we report them in Appendix Section~\ref{baseline-parameter}. 
% The AdamW optimizer is used for all models.
In addition, for VIDRec and VideoRec,  we follow the common practice (e.g., in Video Swin Transformer~\cite{liu2022video}) by selecting a consecutive sequence of five frames from the midsection per video, with a frame interval of 1, to serve as the video input.
\section{Experimental Results}
% In this section, we mainly investigate whether video content helps recommendation
% and how to effectively utilize video contents. IN
The lack of high-quality video datasets has limited research on the effective utilization of raw video content in recommender systems. Here, we provide preliminary exploration with the aim of drawing the community’s attention and inspiring more research on content-driven video recommendation.
\subsection{Benchmark Results \& Analysis}
\label{benchmarkresults}
\begin{table*}[t]
    \caption{Benchmark results on MicroLens-100K. VideoMAE and SlowFast are used as video encoder for VIDRec and VideoRec, respectively (see Footnote\footref{videoencoder}). The fusion of video  and ID embedding features can be achieved through either summation or concatenation, which shows similar results.
    % All hyper-parameters for Method$_{\rm ID}$ and Method$_{\rm ID+V}$ are kept the same for comparison purpose.
    }
    \label{baseline-table}
    \centering
    % \vskip 0.12in
    \scalebox{0.90}{
    \begin{tabular}{llcccc}
        \toprule
        % \multirow{3}{*}{Class} & \multirow{3}{*}{Model} & \multicolumn{4}{c}{MicroLens-100K}\\
        % \cmidrule(lr){3-6} 
        Class & Model & HR@10 & NDCG@10 & HR@20 & NDCG@20  \\
        % & \multicolumn{2}{c}{MIND} & \multicolumn{2}{c}{HM} & \multicolumn{2}{c}{BILI} & \multicolumn{2}{c}{MIND} & \multicolumn{2}{c}{HM} & \multicolumn{2}{c}{BILI} \\
        % \cmidrule(lr){2-3} \cmidrule(lr){4-5} \cmidrule(lr){6-7}\cmidrule(lr){8-9} \cmidrule(lr){10-11} \cmidrule(lr){12-13}
        % & HR@10 & NDCG@10 & HR@20 & NDCG@20 & HR@10 & NDCG@10 & HR@20 & NDCG@20 \\
        % & & H@10 & N@10 & H@20 & N@20 \\
        \midrule
        \multirow{4}{*}{IDRec (CF)} 
        & DSSM~\cite{huang2013learning}      & 0.0394 & 0.0193 & 0.0654 & 0.0258  \\
        & LightGCN~\cite{he2020lightgcn}  & 0.0372 & 0.0177 & 0.0618 & 0.0239 \\
         %& FM~\cite{rendle2010factorization}      &   0.0424     &   0.0208     &   0.0685     &  0.0273     \\
        & NFM~\cite{he2017neuralfm}      &   0.0313     &   0.0159     &   0.0480     &  0.0201     \\
        & DeepFM~\cite{guo2017deepfm}      &   0.0350     &   0.0170    &   0.0571     &  0.0225     \\
\midrule
\multirow{3}{*}{IDRec (SR)}
        & NexItNet~\cite{yuan2019simple} & 0.0805 & 0.0442 & 0.1175 & 0.0535 \\
        & GRU4Rec~\cite{hidasi2015session}  & 0.0782     & 0.0423  & 0.1147 & 0.0515 \\
        & SASRec~\cite{kang2018self}    & 0.0909 & 0.0517 & 0.1278 & 0.0610 \\
\midrule
\multirow{8}{*}{\makecell[l]{VIDRec\\(Frozen Encoder)}} 
        & YouTube$_{\rm ID}$ & 0.0461       & 0.0229       & 0.0747       & 0.0301      \\
        & YouTube$_{\rm ID+V}$~\cite{covington2016deep} &0.0392    &0.0188   & 0.0648   &  0.0252     \\
        & MMGCN$_{\rm ID}$ & 0.0141 & 0.0065 & 0.0247 & 0.0092  \\
        & MMGCN$_{\rm ID+V}$~\cite{wei2019mmgcn} & 0.0214 & 0.0103 & 0.0374 & 0.0143  \\
        & GRCN$_{\rm ID}$ & 0.0282 & 0.0131 & 0.0497 & 0.0185  \\
        & GRCN$_{\rm ID+V}$~\cite{wei2020graph} & 0.0306 & 0.0144 & 0.0547 & 0.0204  \\
        & DSSM$_{\rm ID+V}$ & 0.0279 & 0.0137 & 0.0461 & 0.0183  \\
        & SASRec$_{\rm ID+V}$ & 0.0799 & 0.0415 & 0.1217 & 0.0520  \\
\midrule
        \multirow{3}{*}{\makecell[l]{VideoRec\\(E2E Learning)}}
        & NexItNet$_{\rm V}$~\cite{yuan2019simple} & 0.0862 & 0.0466 & 0.1246 & 0.0562  \\
        & GRU4Rec$_{\rm V}$~\cite{hidasi2015session} & 0.0954 & 0.0517 & 0.1377 & 0.0623 \\
        & SASRec$_{\rm V}$~\cite{kang2018self}  & 0.0948 & 0.0515 & 0.1364 & 0.0619  \\
        % & YouTube$_{\rm V}$~\cite{covington2016deep}  & 0.0395 & 0.0194 & 0.0640 & 0.0256  \\
        \bottomrule
    \end{tabular}
    }
\end{table*}
We evaluate multiple recommender baselines on MicroLens, including IDRec (which does not use video features), VIDRec (which incorporates video features as side information), \& VideoRec (which uses video features exclusively).\footnote{Regarding VIDRec, we extracted video features from VideoMAE~\cite{tong2022videomae}, which demonstrates state-of-the-art (SOTA) accuracy for multiple video understanding tasks (e.g., action classification). For VideoRec, we utilized the SlowFast video network~\cite{feichtenhofer2019slowfast}, which offers the best accuracy through E2E learning. Extensive results on more video encoders are reported Figure~\ref{video-rs}. \label{videoencoder}}
The results are reported in Table~\ref{baseline-table} with the below findings.

\textit{Firstly}, regarding IDRec, all sequential models, including SASRec, NextItNet, and GRU4Rec, outperform non-sequential CF models, namely DSSM, LightGCN, DeepFM, NFM and YouTube.  Among all models, SASRec with Transformer backbone performs the best, improving CNN-based NextItNet and RNN-based GRU4Rec by over 10\%. The findings are consistent with much prior literature~\cite{zhou2020s3,kang2018self,wang2021stackrec,wang2022target}.   
 
\textit{Secondly}, surprisingly, incorporating pre-extracted video features in VIDRec (i.e., GRCN$_{\rm ID+V}$, MMGCN$_{\rm ID+V}$, YouTube$_{\rm ID+V}$, DSSM$_{\rm ID+V}$ and SASRec$_{\rm ID +V}$) does not necessarily result in better performance compared to their IDRec counterparts (e.g.,  YouTube$_{\rm ID}$, DSSM$_{\rm ID}$, and SASRec$_{\rm ID}$). In fact, VIDRec, which treats  video or visual features as side information, is mostly used to assist cold or new item recommendation where pure IDRec is weak due to inadequate training~\cite{he2016vbpr,van2013deep,lee2017large}. However, \textbf{for non-cold or warm item recommendation, such side information may not always improve performance, as ID embeddings may implicitly learn these features.}  Similar findings were also reported in ~\cite{yuan2023go}, which demonstrated that incorporating visual features leads to a decrease in accuracy for non-cold item recommendation. 
These results imply that the common practice of using pre-extracted features from a frozen video encoder may not always yield the expected improvements in performance.

\textit{Thirdly}, the recent study~\cite{yuan2023go} suggested that the optimal way to utilize multimodal features is through E2E training of the recommender model and the item (i.e., video in this case) modality  encoder. Similarly, we observe that VideoRec (i.e., NextItNet$_{\rm V}$, GRU4Rec$_{\rm V}$, and SASRec$_{\rm V}$) achieves the highest recommendation accuracy among all models.
In particular, NextItNet$_{\rm V}$ largely outperforms NextItNet (i.e., NextItNet$_{\rm ID}$), GRU4Rec$_{\rm V}$ largely outperforms GRU4Rec. The comparison clearly demonstrates that learning item representation from raw video data through end-to-end (E2E) training of the video encoder, as opposed to utilizing pre-extracted offline features in VIDRec or pure ID features, leads to superior results (see more analysis in Section~\ref{video-benchmark}). This is likely because E2E training can incorporate both raw video features and collaborative signals from user-item interactions.

% To  the best of our knowledge, We show for the first time that raw video features can replace ID features in regular\footnote{85\% of items have more than 5 user interactions, as shown in Figure~\ref{fig:distribution} (a).} and even warm   (see Appendix Table~\ref{warmup-large-table}) item recommendation settings, rather than just in cold scenarios. We believe this to be a novel contribution.\footnote{This suggests that video features could potentially serve as the primary features, rather than just auxiliary or side information of ID features, in the majority of existing literature.}
Our above findings suggest that \textbf{utilizing raw video features instead of pre-extracted frozen features is crucial for achieving optimal recommendation results, underscoring the significance of the MicroLens video dataset}.

\subsection{Video Understanding Meets Recommender Systems}
\label{video-benchmark}
In the CV field, numerous video networks have been developed. Here,  we aim to explore whether these networks designed for video understanding helps the recommendation task and to what extent.

% Comparing to existing multi-modal recommendation datasets, a distinguishing characteristic of our MicroLens is the extensive volume of raw micro-videos it offers. Capitalizing on this unique feature, we investigate the evolution of video understanding through the lens of human behavior. More precisely, we employ a range of state-of-the-art video classification models as item encoders within the same recommender backbone. By implementing diverse training strategies such as frozen, shallow finetune and full finetune, we endeavor to answer a crucial question: Do superior video models help extract video features more effectively from the standpoint of human comprehension?
Given its top performance in Table~\ref{baseline-table}, we employed SASRec as the recommender backbone and evaluated 15 well-known video encoders that were pre-trained on Kinetics~\cite{kay2017kinetics}, a well-known video (action) classification dataset. These encoders include
R3D-r18~\cite{tran2018closer}, X3D-xs~\cite{feichtenhofer2020x3d}, C2D-r50~\cite{wang2018non}, I3D-r50~\cite{carreira2017quo}, X3D-s~\cite{feichtenhofer2020x3d}, Slow-r50~\cite{fan2021pytorchvideo}, X3D-m~\cite{feichtenhofer2020x3d}, R3D-r50~\cite{tran2018closer}, SlowFast-r50~\cite{feichtenhofer2019slowfast}, CSN-r101~\cite{tran2019video}, X3D-l~\cite{feichtenhofer2020x3d}, SlowFast-r101~\cite{feichtenhofer2019slowfast}, MViT-B-16x4~\cite{fan2021multiscale}, MViT-B-32x3~\cite{fan2021multiscale}, and VideoMAE~\cite{tong2022videomae} with details in Appendix Section~\ref{video-model-description}.
% We report details of them in Appendix Table~\ref{video-model-description}. 
% (sorted in ascending order according to their classification performance on the Kinetics-400 dataset~\cite{kay2017kinetics}).
\begin{figure}[t]
    \centering
    \includegraphics[width=\textwidth]{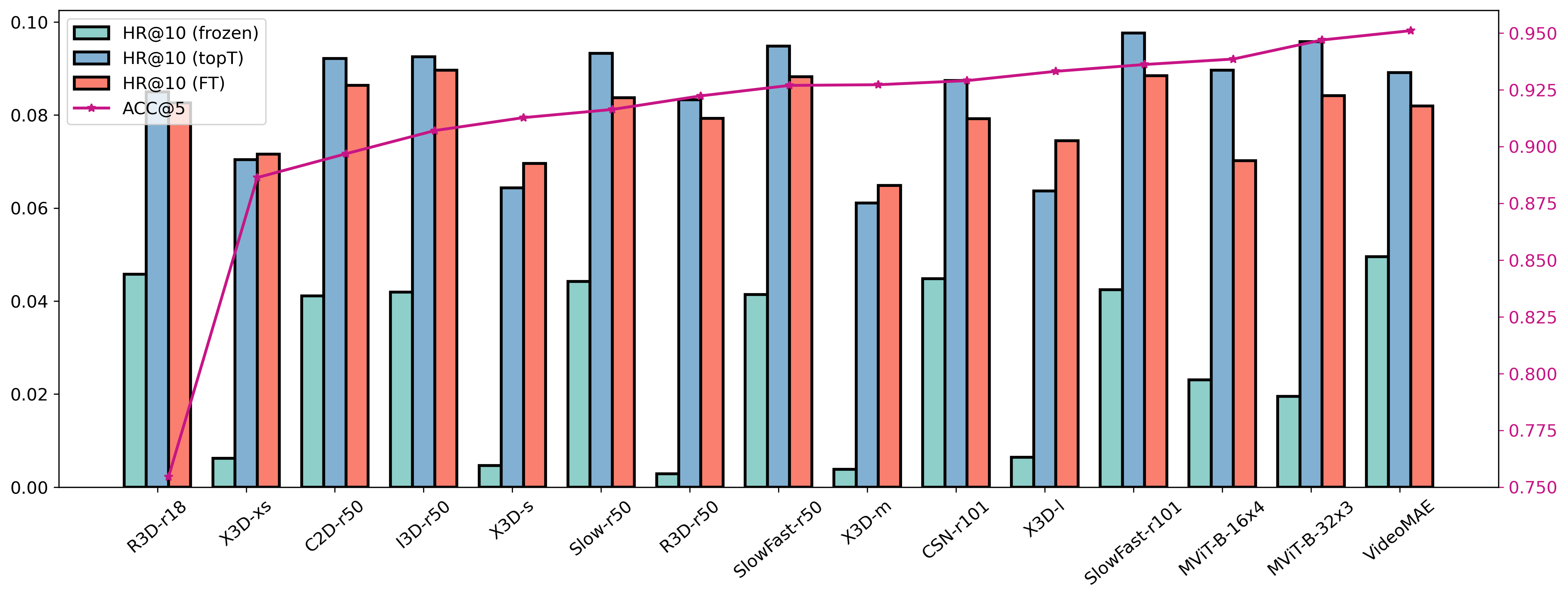}
    \caption{Video recommendation accuracy (bar charts) vs. video classification accuracy (purple line). Frozen means that the video encoder is fixed without parameter update, topT means that only the top few layers of the video encoder are fine-tuned, and FT means  full parameters are fine-tuned.
    % Results of different video models in terms of recommendation and video classification.
    }
    % (size and color to be modified - https://zhuanlan.zhihu.com/p/457797561)
    \label{video-rs}
\end{figure}
% on the architectures, pre-training settings, and corresponding video classification and recommendation performance of these models

\begin{figure}[t]
  \centering
  \subfigure[OT v.s. WT]{
    \begin{minipage}[b]{0.22\textwidth}
    \includegraphics[width=\linewidth]{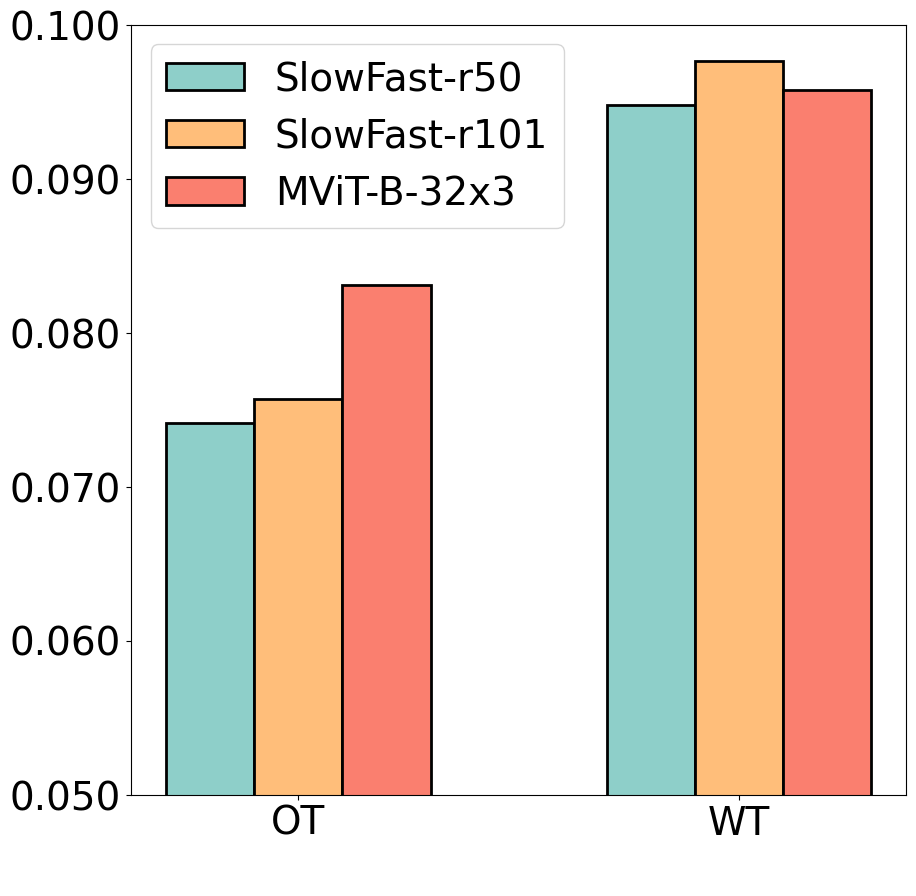}
    
    \end{minipage}
    \label{fig:frz_ft-figure}
  }
    \subfigure[SlowFast-r50]{
    \begin{minipage}[b]{0.22\textwidth}
    \includegraphics[width=\linewidth]{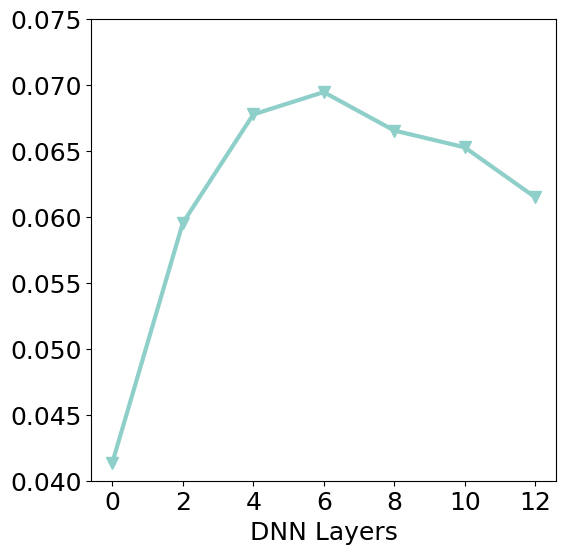}
    
    \end{minipage}
    \label{fig:dnn-sf50-figure}
  }
  % \hfill
  \subfigure[SlowFast-r101]{
    \begin{minipage}[b]{0.22\textwidth}
    \includegraphics[width=\linewidth]{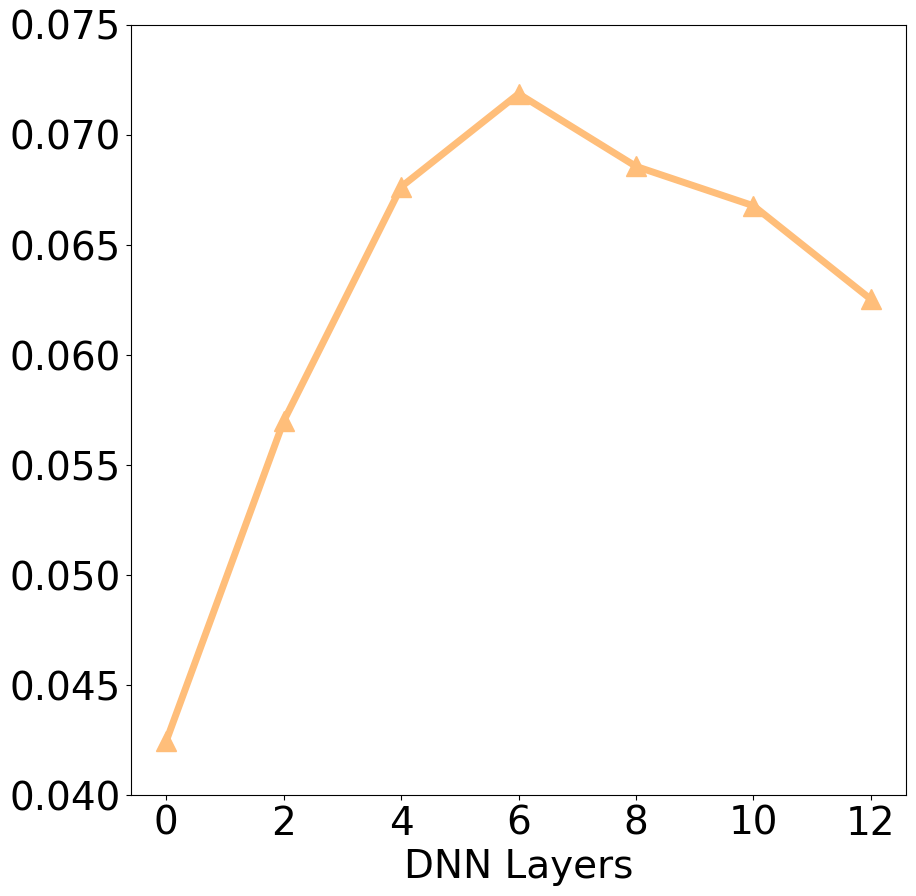}
    
    \end{minipage}
    \label{fig:/dnn-sf101-figure}
  }
  % \hfill
  \subfigure[MVIT-B-32x3]{
    \begin{minipage}[b]{0.22\textwidth}
    \includegraphics[width=\linewidth]{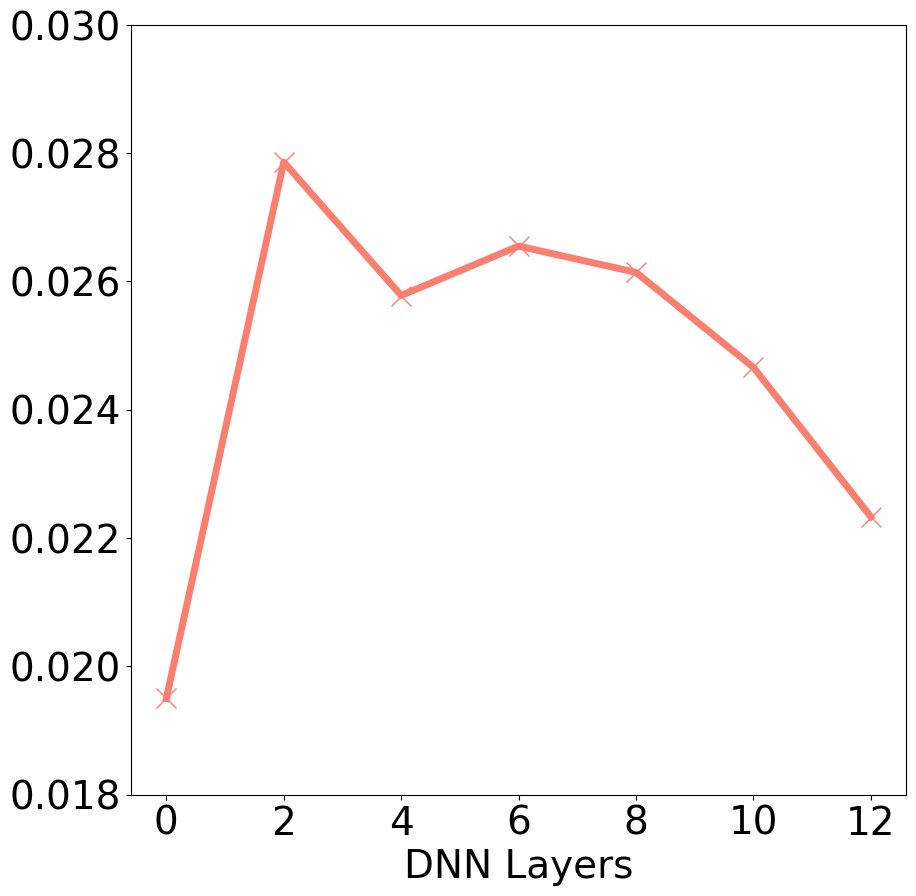}
    
    \end{minipage}
    \label{fig:dnn-mv32-figure}
  }

  \caption{Ablation study of video encoders. (d) "WT" refers to   the video encoders in SASRec$_{\rm V}$ have pre-trained weights from the video classification task, while "OT" denotes that they are randomly initialized. (b) (c) (d) are performance change by adding DNN layers on top of three frozen encoders.
  % Performance of adding DNN layers to SlowFast-r50 (a), SlowFast-r101 (b), MVIT-B-32x3 (c), as well as comparison of full finetune and train from scratch (d). 
  }
  \label{fig:video-exploration}
\end{figure}

\textit{Q1: Can the knowledge learned from video understanding be beneficial for video recommendation?}

Figure~\ref{fig:video-exploration} (a) shows that the recommender model SASRec$_{\rm V}$ with pre-trained SlowFast-r50, SlowFast-r101, and MVIT-B-32x3 video encoders exhibited a solid improvement in performance compared to their random initialization versions. These results clearly suggest that \textbf{the parameters learned from the video understanding task  in the CV field are highly valuable for improving video recommendations.} 

\textit{Q2: Does a strong video encoder always  translate into a better video recommender model?}

Figure~\ref{video-rs} compares the video classification (VC) task and the video recommendation task. It is evident  that \textbf{a higher performance in the VC task (purple line) does not necessarily correspond to a higher accuracy in video recommendation (bar charts).} For instance, while VideoMAE achieves optimal results in video classification, it does not necessarily guarantee the highest accuracy for item recommendation. A more pronounced example is R3D-r18, which exhibits the worst results in video classification but performs relatively well in the recommendation task.
This finding differs from~\cite{yuan2023go}, which demonstrated that higher performance in NLP and CV models generally leads to higher recommendation accuracy. However, it should be noted that~\cite{yuan2023go} only investigated image and text recommendation, and did not explore video recommendation, which could be very different.

\textit{Q3: Are the semantic representations learned from the video understanding task universal  for video recommendation}

In Section~\ref{benchmarkresults}, we showed that incorporating pre-extracted video features may not necessarily improve recommendation accuracy when ID features are sufficiently trained. Here, we conducted a more detailed study by comparing the performance of recommender models (i.e.,  SASRec) with frozen (equivalent to pre-extracted video features) and end-to-end trained video encoders. Figure~\ref{video-rs} clearly demonstrates that SASRec$_{\rm V}$ with retrained video encoders, whether topT or FT, performs significantly better, with about a 2-fold improvement over the frozen approach. These results suggest that \textbf{the video semantic representations learned by the popular video classification task are not universal to the recommendation task, and retraining the video encoder on the recommendation data is necessary to achieve optimal performance.} This is because, if the pre-extracted video features were a perfect representation, a linear layer applied to these features is enough to perform equally well as the fine-tuned video encoder. Although adding more DNN layers on the pre-extracted video features significantly improves accuracy (see Figure~\ref{fig:video-exploration}(b,c,d)), it still largely falls  short of the accuracy achieved by using a fine-tuned video encoder.
Moreover, the results indicate that full parameter fine-tuning  (FT) of the video encoder is not necessary, as fine-tuning only the top few layers (TopT) generally produces superior results. This seems reasonable since optimizing all parameters of the video encoder may result in complete  catastrophic forgetting of the knowledge learned during the video pre-training task. This  highlights once again  the value of the knowledge gained from video understanding tasks for video recommendation.

To sum up, existing video understanding technologies, including video encoders and trained parameters,  are undoubtedly valuable for video recommendation. However, there is still a significant semantic gap between video understanding tasks and recommendation systems. Therefore, not all advances made in video tasks can directly translate into improvements for recommender systems.

\subsection{Additional Exploration of VideoRec}
Beyond the above results, we have performed other interesting empirical experiments as below.

\textit{Q1: How would the recommendation performance be impacted if we solely rely on the cover image instead of the raw video? }

% For image baselines, we take the best architecture SASRec as the backbone and apply different SOTA image encoders, i.e., ResNet (r18), Swin (tiny), and MAE (base). (exhaustive search for hyper-parameters)
To answer this question, we use three SOTA image encoders to represent video cover images. We still use the E2E learning and refer to this approach as ImageRec.
% We conducte an exhaustive search for hyper-parameters for three recommendation models: SASRec${\rm MAE}$, NextItNet${\rm MAE}$, and GRU4Rec$_{\rm MAE}$.
Our results are given in Table~\ref{image-baselines}, which suggests that VideoRec generally outperforms ImageRec when compared to the results of SASRec$_{\rm V}$ in Table~\ref{baseline-table} and Figure~\ref{video-rs}. This also reflects the importance of video content for recommender systems.

% used the SOTA MAE image encoder~\cite{he2022masked} to represent videos by their cover images and perform E2E learning, which we refer to as ImageRec. 
\textit{Q2: Can VideoRec compete with IDRec in recommending highly popular items?}

In Section~\ref{benchmarkresults}, we showed that VideoRec is capable of surpassing IDRec in the regular item  recommendation setting (including both popular and cold items). Here, we want to further investigate whether VideoRec still outperforms IDRec in recommending popular items. The reason why we are keen in comparing IDRec is that many recent studies~\cite{yuan2023go,li2023exploring,wang2022transrec,li2023text,hou2023learning,fu2023exploring} have claimed that IDRec poses a major obstacle for \textit{transferable} or \textit{foundation} recommender models~\cite{geng2023vip5} as ID features are generally non-shareable in practice. Appendix Table~\ref{warmup-table} shows that VideoRec using the SOTA SASRec architecture can consistently outperform IDRec, even in very warm item settings.

% IDRec is widely recognized as the leading baseline for recommending warm or popular items.  Besides, to our best knowledge, there is currently no literature indicating that using only video content can outperform IDRec in recommending popular items.
To  our best knowledge, \textbf{this study is the first to show that raw video features can potentially replace ID features in both  \textit{warm} and cold\footnote{More improvements can be easily observed on cold items (see Appendix Figure~\ref{cold-start}), which was also studied in much prior literature~\cite{kumar2018icebreaker,lee2017large,yuan2023go}.} item recommendation settings.} We consider this to be a significant contribution as it suggests that VideoRec may potentially challenge the dominant role of ID-based recommender systems. This is particularly noteworthy given VideoRec's natural advantage in transfer learning due to the generality of video or visual features. That is, VideoRec has taken a key step towards  the grand  goal of a universal "one-for-all"  recommender paradigm.

% akin to  BERT~\cite{devlin2018bert} and ChatGPT~\cite{zhang2023complete} in the NLP field.
% \footnote{This suggests that video features could potentially serve as the primary features, rather than just auxiliary or side information of ID features, in the majority of existing literature.}

At last, we have reported some key baseline results in MicroLens-1M in Appendix Table~\ref{baseline-large-table} and~\ref{warmup-large-table}.
\begin{table*}[t]
    \caption{Recommendation accuracy using cover images to represent videos, with three SOTA image encoders, i.e., 
    % ResNet (r18)~\cite{he2016deep}, Swin Transformer (tiny)~\cite{liu2021swin} and MAE (base)~\cite{he2022masked} (see Appendix Section~\ref{Cover_vs_video} for details).
        ResNet~\cite{he2016deep}, Swin Transformer~\cite{liu2021swin} and MAE~\cite{he2022masked} (see Appendix Section~\ref{MEversion} for details).
    }
    \label{image-baselines}
    \centering
    % \vskip 0.12in
    \scalebox{0.90}{
    \begin{tabular}{lcccc}
        \toprule
        \multirow{3}{*}{Model} & \multicolumn{4}{c}{MicroLens-100K}  \\
        \cmidrule(lr){2-5} 
        % & \multicolumn{2}{c}{MIND} & \multicolumn{2}{c}{HM} & \multicolumn{2}{c}{BILI} & \multicolumn{2}{c}{MIND} & \multicolumn{2}{c}{HM} & \multicolumn{2}{c}{BILI} \\
        % \cmidrule(lr){2-3} \cmidrule(lr){4-5} \cmidrule(lr){6-7}\cmidrule(lr){8-9} \cmidrule(lr){10-11} \cmidrule(lr){12-13}
        % & HR@10 & NDCG@10 & HR@20 & NDCG@20 & HR@10 & NDCG@10 & HR@20 & NDCG@20 \\
        & HR@10 & NDCG@10 & HR@20 & NDCG@20\\
        \midrule
        % VBPR      & 0.0344 & 0.0169 & 0.0554 & 0.0222 & 0.0109& 0.0053 & 0.0187 & 0.0072 \\
        % VisRank & 0.0230 & 0.0123 & 0.0342 & 0.0151 & 0.0095 & 0.0050 & 0.0145 & 0.0063 \\
        SASRec$_{\rm ResNet}$ &  0.0858 & 0.0462 & 0.1264 & 0.0564 \\
        SASRec$_{\rm MAE}$ & 0.0828 & 0.0447 & 0.1223 & 0.0546 \\
        SASRec$_{\rm Swin}$ & 0.0892 & 0.0479 & 0.1299 & 0.0582  \\
        \bottomrule
    \end{tabular}
    }
\end{table*}

\section{Conclusions and Broader Impact}
This paper introduces ``\textit{MicroLens}'', the most immense and diverse micro-video dataset to date. Each video in MicroLens contains rich modalities, including text descriptions, images, audio, and raw video information. We conduct an extensive empirical study and benchmark multiple classical recommender baselines. The newly proposed method, VideoRec, directly learns item representations from raw video features and achieves the highest recommendation accuracy among the compared models. We anticipate that MicroLens will become a valuable resource for the
recommender system community, enabling multiple research directions in multimodal or micro-video recommendation.

Although MicroLens is primarily used for video recommendation tasks in this paper, there are other important research directions worth exploring. For instance, recent advances in foundation one-for-all models, such as ChatGPT~\cite{ouyang2022training} and GPT-4~\cite{openai2023gpt4}, have achieved remarkable success in the fields of NLP and CV. However, the recommender system community has made limited progress in large foundation models, particularly in vision- or video-content driven recommender systems. This is partly due to the lack of large-scale, diverse, and high-quality multimodal recommendation datasets, which presents a significant challenge. We envision that MicroLens may serve as a valuable pre-training dataset for visually relevant recommendation, as a single micro-video in MicroLens can generate hundreds of high-quality images,  resulting in a trillion-level of user-image interactions.

Moreover, the field of video understanding has recently made significant strides and is poised to become a future research hotspot~\cite{bertasius2021space,arnab2021vivit,tong2022videomae,tran2018closer}. Using video understanding to drive more fine-grained recommendation, rather than simply learning user behavior similarities, is undoubtedly a more promising direction. Additionally, treating video recommendation as a downstream task for video understanding has the potential to unite the two communities and foster mutual development.

\clearpage

\appendix

\bibliographystyle{plain}
\bibliography{arxiv_2023}

\begin{thebibliography}{10}

\bibitem{abu2016youtube}
Sami Abu-El-Haija, Nisarg Kothari, Joonseok Lee, Paul Natsev, George Toderici, Balakrishnan Varadarajan, and Sudheendra Vijayanarasimhan.
\newblock Youtube-8m: A large-scale video classification benchmark.
\newblock {\em arXiv preprint arXiv:1609.08675}, 2016.

\bibitem{arnab2021vivit}
Anurag Arnab, Mostafa Dehghani, Georg Heigold, Chen Sun, Mario Lu{\v{c}}i{\'c}, and Cordelia Schmid.
\newblock Vivit: A video vision transformer.
\newblock In {\em Proceedings of the IEEE/CVF international conference on computer vision}, pages 6836--6846, 2021.

\bibitem{bertasius2021space}
Gedas Bertasius, Heng Wang, and Lorenzo Torresani.
\newblock Is space-time attention all you need for video understanding?
\newblock In {\em ICML}, volume~2, page~4, 2021.

\bibitem{carreira2017quo}
Joao Carreira and Andrew Zisserman.
\newblock Quo vadis, action recognition? a new model and the kinetics dataset.
\newblock In {\em proceedings of the IEEE Conference on Computer Vision and Pattern Recognition}, pages 6299--6308, 2017.

\bibitem{chen2019pog}
Wen Chen, Pipei Huang, Jiaming Xu, Xin Guo, Cheng Guo, Fei Sun, Chao Li, Andreas Pfadler, Huan Zhao, and Binqiang Zhao.
\newblock Pog: personalized outfit generation for fashion recommendation at alibaba ifashion.
\newblock In {\em Proceedings of the 25th ACM SIGKDD international conference on knowledge discovery \& data mining}, pages 2662--2670, 2019.

\bibitem{chen2023reasoner}
Xu~Chen, Jingsen Zhang, Lei Wang, Quanyu Dai, Zhenhua Dong, Ruiming Tang, Rui Zhang, Li~Chen, and Ji-Rong Wen.
\newblock Reasoner: An explainable recommendation dataset with multi-aspect real user labeled ground truths towards more measurable explainable recommendation.
\newblock {\em arXiv preprint arXiv:2303.00168}, 2023.

\bibitem{covington2016deep}
Paul Covington, Jay Adams, and Emre Sargin.
\newblock Deep neural networks for youtube recommendations.
\newblock In {\em Proceedings of the 10th ACM conference on recommender systems}, pages 191--198, 2016.

\bibitem{fan2021pytorchvideo}
Haoqi Fan, Tullie Murrell, Heng Wang, Kalyan~Vasudev Alwala, Yanghao Li, Yilei Li, Bo~Xiong, Nikhila Ravi, Meng Li, Haichuan Yang, et~al.
\newblock Pytorchvideo: A deep learning library for video understanding.
\newblock In {\em Proceedings of the 29th ACM International Conference on Multimedia}, pages 3783--3786, 2021.

\bibitem{fan2021multiscale}
Haoqi Fan, Bo~Xiong, Karttikeya Mangalam, Yanghao Li, Zhicheng Yan, Jitendra Malik, and Christoph Feichtenhofer.
\newblock Multiscale vision transformers.
\newblock In {\em Proceedings of the IEEE/CVF International Conference on Computer Vision}, pages 6824--6835, 2021.

\bibitem{feichtenhofer2020x3d}
Christoph Feichtenhofer.
\newblock X3d: Expanding architectures for efficient video recognition.
\newblock In {\em Proceedings of the IEEE/CVF conference on computer vision and pattern recognition}, pages 203--213, 2020.

\bibitem{feichtenhofer2019slowfast}
Christoph Feichtenhofer, Haoqi Fan, Jitendra Malik, and Kaiming He.
\newblock Slowfast networks for video recognition.
\newblock In {\em Proceedings of the IEEE/CVF international conference on computer vision}, pages 6202--6211, 2019.

\bibitem{fu2023exploring}
Junchen Fu, Fajie Yuan, Yu~Song, Zheng Yuan, Mingyue Cheng, Shenghui Cheng, Jiaqi Zhang, Jie Wang, and Yunzhu Pan.
\newblock Exploring adapter-based transfer learning for recommender systems: Empirical studies and practical insights.
\newblock {\em arXiv preprint arXiv:2305.15036}, 2023.

\bibitem{gao2022kuairec}
Chongming Gao, Shijun Li, Wenqiang Lei, Jiawei Chen, Biao Li, Peng Jiang, Xiangnan He, Jiaxin Mao, and Tat-Seng Chua.
\newblock Kuairec: A fully-observed dataset and insights for evaluating recommender systems.
\newblock In {\em Proceedings of the 31st ACM International Conference on Information \& Knowledge Management}, pages 540--550, 2022.

\bibitem{gao2022kuairand}
Chongming Gao, Shijun Li, Yuan Zhang, Jiawei Chen, Biao Li, Wenqiang Lei, Peng Jiang, and Xiangnan He.
\newblock Kuairand: An unbiased sequential recommendation dataset with randomly exposed videos.
\newblock In {\em Proceedings of the 31st ACM International Conference on Information and Knowledge Management}, CIKM '22, page 3953–3957, 2022.

\bibitem{geng2023vip5}
Shijie Geng, Juntao Tan, Shuchang Liu, Zuohui Fu, and Yongfeng Zhang.
\newblock Vip5: Towards multimodal foundation models for recommendation.
\newblock {\em arXiv preprint arXiv:2305.14302}, 2023.

\bibitem{gong2022real}
Xudong Gong, Qinlin Feng, Yuan Zhang, Jiangling Qin, Weijie Ding, Biao Li, Peng Jiang, and Kun Gai.
\newblock Real-time short video recommendation on mobile devices.
\newblock In {\em Proceedings of the 31st ACM International Conference on Information \& Knowledge Management}, pages 3103--3112, 2022.

\bibitem{guo2017deepfm}
Huifeng Guo, Ruiming Tang, Yunming Ye, Zhenguo Li, and Xiuqiang He.
\newblock Deepfm: a factorization-machine based neural network for ctr prediction.
\newblock {\em arXiv preprint arXiv:1703.04247}, 2017.

\bibitem{han2016dancelets}
Tingting Han, Hongxun Yao, Chenliang Xu, Xiaoshuai Sun, Yanhao Zhang, and Jason~J Corso.
\newblock Dancelets mining for video recommendation based on dance styles.
\newblock {\em IEEE Transactions on Multimedia}, 19(4):712--724, 2016.

\bibitem{harper2015movielens}
F~Maxwell Harper and Joseph~A Konstan.
\newblock The movielens datasets: History and context.
\newblock {\em Acm transactions on interactive intelligent systems (tiis)}, 5(4):1--19, 2015.

\bibitem{he2022masked}
Kaiming He, Xinlei Chen, Saining Xie, Yanghao Li, Piotr Doll{\'a}r, and Ross Girshick.
\newblock Masked autoencoders are scalable vision learners.
\newblock In {\em Proceedings of the IEEE/CVF Conference on Computer Vision and Pattern Recognition}, pages 16000--16009, 2022.

\bibitem{he2016deep}
Kaiming He, Xiangyu Zhang, Shaoqing Ren, and Jian Sun.
\newblock Deep residual learning for image recognition.
\newblock In {\em Proceedings of the IEEE conference on computer vision and pattern recognition}, pages 770--778, 2016.

\bibitem{he2016vista}
Ruining He, Chen Fang, Zhaowen Wang, and Julian McAuley.
\newblock Vista: A visually, socially, and temporally-aware model for artistic recommendation.
\newblock In {\em Proceedings of the 10th ACM conference on recommender systems}, pages 309--316, 2016.

\bibitem{he2016ups}
Ruining He and Julian McAuley.
\newblock Ups and downs: Modeling the visual evolution of fashion trends with one-class collaborative filtering.
\newblock In {\em proceedings of the 25th international conference on world wide web}, pages 507--517, 2016.

\bibitem{he2016vbpr}
Ruining He and Julian McAuley.
\newblock Vbpr: visual bayesian personalized ranking from implicit feedback.
\newblock In {\em Proceedings of the AAAI conference on artificial intelligence}, volume~30, 2016.

\bibitem{he2017neuralfm}
Xiangnan He and Tat-Seng Chua.
\newblock Neural factorization machines for sparse predictive analytics.
\newblock In {\em Proceedings of the 40th International ACM SIGIR conference on Research and Development in Information Retrieval}, pages 355--364, 2017.

\bibitem{he2020lightgcn}
Xiangnan He, Kuan Deng, Xiang Wang, Yan Li, Yongdong Zhang, and Meng Wang.
\newblock Lightgcn: Simplifying and powering graph convolution network for recommendation.
\newblock In {\em Proceedings of the 43rd International ACM SIGIR conference on research and development in Information Retrieval}, pages 639--648, 2020.

\bibitem{hidasi2015session}
Bal{\'a}zs Hidasi, Alexandros Karatzoglou, Linas Baltrunas, and Domonkos Tikk.
\newblock Session-based recommendations with recurrent neural networks.
\newblock {\em arXiv preprint arXiv:1511.06939}, 2015.

\bibitem{hou2023learning}
Yupeng Hou, Zhankui He, Julian McAuley, and Wayne~Xin Zhao.
\newblock Learning vector-quantized item representation for transferable sequential recommenders.
\newblock In {\em Proceedings of the ACM Web Conference 2023}, pages 1162--1171, 2023.

\bibitem{huang2013learning}
Po-Sen Huang, Xiaodong He, Jianfeng Gao, Li~Deng, Alex Acero, and Larry Heck.
\newblock Learning deep structured semantic models for web search using clickthrough data.
\newblock In {\em Proceedings of the 22nd ACM international conference on Information \& Knowledge Management}, pages 2333--2338, 2013.

\bibitem{jiang2020aspect}
Hao Jiang, Wenjie Wang, Yinwei Wei, Zan Gao, Yinglong Wang, and Liqiang Nie.
\newblock What aspect do you like: Multi-scale time-aware user interest modeling for micro-video recommendation.
\newblock In {\em Proceedings of the 28th ACM International conference on Multimedia}, pages 3487--3495, 2020.

\bibitem{kang2018self}
Wang-Cheng Kang and Julian McAuley.
\newblock Self-attentive sequential recommendation.
\newblock In {\em 2018 IEEE international conference on data mining (ICDM)}, pages 197--206. IEEE, 2018.

\bibitem{kay2017kinetics}
Will Kay, Joao Carreira, Karen Simonyan, Brian Zhang, Chloe Hillier, Sudheendra Vijayanarasimhan, Fabio Viola, Tim Green, Trevor Back, Paul Natsev, et~al.
\newblock The kinetics human action video dataset.
\newblock {\em arXiv preprint arXiv:1705.06950}, 2017.

\bibitem{kumar2018icebreaker}
Yaman Kumar, Agniv Sharma, Abhigyan Khaund, Akash Kumar, Ponnurangam Kumaraguru, Rajiv~Ratn Shah, and Roger Zimmermann.
\newblock Icebreaker: Solving cold start problem for video recommendation engines.
\newblock In {\em 2018 IEEE international symposium on multimedia (ISM)}, pages 217--222. IEEE, 2018.

\bibitem{lee2017large}
Joonseok Lee and Sami Abu-El-Haija.
\newblock Large-scale content-only video recommendation.
\newblock In {\em Proceedings of the IEEE International Conference on Computer Vision Workshops}, pages 987--995, 2017.

\bibitem{lei2021semi}
Chenyi Lei, Yong Liu, Lingzi Zhang, Guoxin Wang, Haihong Tang, Houqiang Li, and Chunyan Miao.
\newblock Semi: A sequential multi-modal information transfer network for e-commerce micro-video recommendations.
\newblock In {\em Proceedings of the 27th ACM SIGKDD Conference on Knowledge Discovery \& Data Mining}, pages 3161--3171, 2021.

\bibitem{li2023text}
Jiacheng Li, Ming Wang, Jin Li, Jinmiao Fu, Xin Shen, Jingbo Shang, and Julian McAuley.
\newblock Text is all you need: Learning language representations for sequential recommendation.
\newblock {\em arXiv preprint arXiv:2305.13731}, 2023.

\bibitem{li2023exploring}
Ruyu Li, Wenhao Deng, Yu~Cheng, Zheng Yuan, Jiaqi Zhang, and Fajie Yuan.
\newblock Exploring the upper limits of text-based collaborative filtering using large language models: Discoveries and insights.
\newblock {\em arXiv preprint arXiv:2305.11700}, 2023.

\bibitem{liu2019user}
Shang Liu, Zhenzhong Chen, Hongyi Liu, and Xinghai Hu.
\newblock User-video co-attention network for personalized micro-video recommendation.
\newblock In {\em The World Wide Web Conference}, pages 3020--3026, 2019.

\bibitem{liu2021concept}
Yiyu Liu, Qian Liu, Yu~Tian, Changping Wang, Yanan Niu, Yang Song, and Chenliang Li.
\newblock Concept-aware denoising graph neural network for micro-video recommendation.
\newblock In {\em Proceedings of the 30th ACM International Conference on Information \& Knowledge Management}, pages 1099--1108, 2021.

\bibitem{liu2021swin}
Ze~Liu, Yutong Lin, Yue Cao, Han Hu, Yixuan Wei, Zheng Zhang, Stephen Lin, and Baining Guo.
\newblock Swin transformer: Hierarchical vision transformer using shifted windows.
\newblock In {\em Proceedings of the IEEE/CVF international conference on computer vision}, pages 10012--10022, 2021.

\bibitem{liu2022video}
Ze~Liu, Jia Ning, Yue Cao, Yixuan Wei, Zheng Zhang, Stephen Lin, and Han Hu.
\newblock Video swin transformer.
\newblock In {\em Proceedings of the IEEE/CVF conference on computer vision and pattern recognition}, pages 3202--3211, 2022.

\bibitem{nielsen2022mumin}
Dan~S Nielsen and Ryan McConville.
\newblock Mumin: A large-scale multilingual multimodal fact-checked misinformation social network dataset.
\newblock In {\em Proceedings of the 45th International ACM SIGIR Conference on Research and Development in Information Retrieval}, pages 3141--3153, 2022.

\bibitem{openai2023gpt4}
OpenAI.
\newblock Gpt-4 technical report, 2023.

\bibitem{ouyang2022training}
Long Ouyang, Jeffrey Wu, Xu~Jiang, Diogo Almeida, Carroll Wainwright, Pamela Mishkin, Chong Zhang, Sandhini Agarwal, Katarina Slama, Alex Ray, et~al.
\newblock Training language models to follow instructions with human feedback.
\newblock {\em Advances in Neural Information Processing Systems}, 35:27730--27744, 2022.

\bibitem{tong2022videomae}
Zhan Tong, Yibing Song, Jue Wang, and Limin Wang.
\newblock Videomae: Masked autoencoders are data-efficient learners for self-supervised video pre-training.
\newblock {\em arXiv preprint arXiv:2203.12602}, 2022.

\bibitem{tran2019video}
Du~Tran, Heng Wang, Lorenzo Torresani, and Matt Feiszli.
\newblock Video classification with channel-separated convolutional networks.
\newblock In {\em Proceedings of the IEEE/CVF International Conference on Computer Vision}, pages 5552--5561, 2019.

\bibitem{tran2018closer}
Du~Tran, Heng Wang, Lorenzo Torresani, Jamie Ray, Yann LeCun, and Manohar Paluri.
\newblock A closer look at spatiotemporal convolutions for action recognition.
\newblock In {\em Proceedings of the IEEE conference on Computer Vision and Pattern Recognition}, pages 6450--6459, 2018.

\bibitem{van2013deep}
Aaron Van~den Oord, Sander Dieleman, and Benjamin Schrauwen.
\newblock Deep content-based music recommendation.
\newblock {\em Advances in neural information processing systems}, 26, 2013.

\bibitem{wang2022target}
Chenyang Wang, Zhefan Wang, Yankai Liu, Yang Ge, Weizhi Ma, Min Zhang, Yiqun Liu, Junlan Feng, Chao Deng, and Shaoping Ma.
\newblock Target interest distillation for multi-interest recommendation.
\newblock In {\em Proceedings of the 31st ACM International Conference on Information \& Knowledge Management}, pages 2007--2016, 2022.

\bibitem{wang2021stackrec}
Jiachun Wang, Fajie Yuan, Jian Chen, Qingyao Wu, Min Yang, Yang Sun, and Guoxiao Zhang.
\newblock Stackrec: Efficient training of very deep sequential recommender models by iterative stacking.
\newblock In {\em Proceedings of the 44th International ACM SIGIR conference on Research and Development in Information Retrieval}, pages 357--366, 2021.

\bibitem{wang2022transrec}
Jie Wang, Fajie Yuan, Mingyue Cheng, Joemon~M Jose, Chenyun Yu, Beibei Kong, Zhijin Wang, Bo~Hu, and Zang Li.
\newblock Transrec: Learning transferable recommendation from mixture-of-modality feedback.
\newblock {\em arXiv preprint arXiv:2206.06190}, 2022.

\bibitem{wang2018non}
Xiaolong Wang, Ross Girshick, Abhinav Gupta, and Kaiming He.
\newblock Non-local neural networks.
\newblock In {\em Proceedings of the IEEE conference on computer vision and pattern recognition}, pages 7794--7803, 2018.

\bibitem{wei2020graph}
Yinwei Wei, Xiang Wang, Liqiang Nie, Xiangnan He, and Tat-Seng Chua.
\newblock Graph-refined convolutional network for multimedia recommendation with implicit feedback.
\newblock In {\em Proceedings of the 28th ACM international conference on multimedia}, pages 3541--3549, 2020.

\bibitem{wei2019mmgcn}
Yinwei Wei, Xiang Wang, Liqiang Nie, Xiangnan He, Richang Hong, and Tat-Seng Chua.
\newblock Mmgcn: Multi-modal graph convolution network for personalized recommendation of micro-video.
\newblock In {\em Proceedings of the 27th ACM international conference on multimedia}, pages 1437--1445, 2019.

\bibitem{woolridge2021sequence}
Daniel Woolridge, Sean Wilner, and Madeleine Glick.
\newblock Sequence or pseudo-sequence? an analysis of sequential recommendation datasets.
\newblock In {\em Perspectives@ RecSys}, 2021.

\bibitem{wu2020mind}
Fangzhao Wu, Ying Qiao, Jiun-Hung Chen, Chuhan Wu, Tao Qi, Jianxun Lian, Danyang Liu, Xing Xie, Jianfeng Gao, Winnie Wu, et~al.
\newblock Mind: A large-scale dataset for news recommendation.
\newblock In {\em Proceedings of the 58th Annual Meeting of the Association for Computational Linguistics}, pages 3597--3606, 2020.

\bibitem{wu2019hierarchical}
Le~Wu, Lei Chen, Richang Hong, Yanjie Fu, Xing Xie, and Meng Wang.
\newblock A hierarchical attention model for social contextual image recommendation.
\newblock {\em IEEE Transactions on Knowledge and Data Engineering}, 32(10):1854--1867, 2019.

\bibitem{yan2022personalized}
An~Yan, Zhankui He, Jiacheng Li, Tianyang Zhang, and Julian McAuley.
\newblock Personalized showcases: Generating multi-modal explanations for recommendations.
\newblock {\em arXiv preprint arXiv:2207.00422}, 2022.

\bibitem{yang2022gram}
Yoonseok Yang, Kyu~Seok Kim, Minsam Kim, and Juneyoung Park.
\newblock Gram: Fast fine-tuning of pre-trained language models for content-based collaborative filtering.
\newblock {\em arXiv preprint arXiv:2204.04179}, 2022.

\bibitem{yi2019sampling}
Xinyang Yi, Ji~Yang, Lichan Hong, Derek~Zhiyuan Cheng, Lukasz Heldt, Aditee Kumthekar, Zhe Zhao, Li~Wei, and Ed~Chi.
\newblock Sampling-bias-corrected neural modeling for large corpus item recommendations.
\newblock In {\em Proceedings of the 13th ACM Conference on Recommender Systems}, pages 269--277, 2019.

\bibitem{yu2023improving}
Yisong Yu, Beihong Jin, Jiageng Song, Beibei Li, Yiyuan Zheng, and Wei Zhuo.
\newblock Improving micro-video recommendation by controlling position bias.
\newblock In {\em Machine Learning and Knowledge Discovery in Databases: European Conference, ECML PKDD 2022, Grenoble, France, September 19--23, 2022, Proceedings, Part I}, pages 508--523. Springer, 2023.

\bibitem{yuan2019simple}
Fajie Yuan, Alexandros Karatzoglou, Ioannis Arapakis, Joemon~M Jose, and Xiangnan He.
\newblock A simple convolutional generative network for next item recommendation.
\newblock In {\em Proceedings of the twelfth ACM international conference on web search and data mining}, pages 582--590, 2019.

\bibitem{yuan2022tenrec}
Guanghu Yuan, Fajie Yuan, Yudong Li, Beibei Kong, Shujie Li, Lei Chen, Min Yang, Chenyun Yu, Bo~Hu, Zang Li, et~al.
\newblock Tenrec: A large-scale multipurpose benchmark dataset for recommender systems.
\newblock {\em arXiv preprint arXiv:2210.10629}, 2022.

\bibitem{yuan2023go}
Zheng Yuan, Fajie Yuan, Yu~Song, Youhua Li, Junchen Fu, Fei Yang, Yunzhu Pan, and Yongxin Ni.
\newblock Where to go next for recommender systems? id-vs. modality-based recommender models revisited.
\newblock {\em arXiv preprint arXiv:2303.13835}, 2023.

\bibitem{zeng2022tencent}
Zhaoyang Zeng, Yongsheng Luo, Zhenhua Liu, Fengyun Rao, Dian Li, Weidong Guo, and Zhen Wen.
\newblock Tencent-mvse: A large-scale benchmark dataset for multi-modal video similarity evaluation.
\newblock In {\em Proceedings of the IEEE/CVF Conference on Computer Vision and Pattern Recognition}, pages 3138--3147, 2022.

\bibitem{zheng2022dvr}
Yu~Zheng, Chen Gao, Jingtao Ding, Lingling Yi, Depeng Jin, Yong Li, and Meng Wang.
\newblock Dvr: Micro-video recommendation optimizing watch-time-gain under duration bias.
\newblock In {\em Proceedings of the 30th ACM International Conference on Multimedia}, pages 334--345, 2022.

\bibitem{zhou2018deep}
Guorui Zhou, Xiaoqiang Zhu, Chenru Song, Ying Fan, Han Zhu, Xiao Ma, Yanghui Yan, Junqi Jin, Han Li, and Kun Gai.
\newblock Deep interest network for click-through rate prediction.
\newblock In {\em Proceedings of the 24th ACM SIGKDD International Conference on Knowledge Discovery \& Data Mining}, pages 1059--1068, 2018.

\bibitem{zhou2020s3}
Kun Zhou, Hui Wang, Wayne~Xin Zhao, Yutao Zhu, Sirui Wang, Fuzheng Zhang, Zhongyuan Wang, and Ji-Rong Wen.
\newblock S3-rec: Self-supervised learning for sequential recommendation with mutual information maximization.
\newblock In {\em Proceedings of the 29th ACM international conference on information \& knowledge management}, pages 1893--1902, 2020.

\bibitem{zhu2018learning}
Han Zhu, Xiang Li, Pengye Zhang, Guozheng Li, Jie He, Han Li, and Kun Gai.
\newblock Learning tree-based deep model for recommender systems.
\newblock In {\em Proceedings of the 24th ACM SIGKDD International Conference on Knowledge Discovery \& Data Mining}, pages 1079--1088, 2018.

\end{thebibliography}

%%%%%%%%%%%%%%%%%%%%%%%%%%%%%%%%%%%%%%%%%%%%%%%%%%%%%%%%%%%%

\clearpage

\section{Technical Details for Data Integration}
\label{technical-details}
When a collection node obtained download links, it was responsible for distributing these links to the download nodes. Multiple download nodes were utilized, each equipped with large-scale storage and high-speed broadband. The download nodes were able to communicate and collaborate with each other to ensure efficient and non-redundant downloading of images, audio, and video files. Upon the completion of the downloading process, multiple high-speed transfer channels were established between the download nodes and the data integration node. This allowed for the merging of all downloaded files into a single root directory. The data integration node utilized a high-capacity hard drive to store all the downloaded data. The data integration process allowed us to effectively manage and organize a large amount of collected data, enabling us to analyze and extract valuable insights from the data. 

 Overall, our data integration process allowed us to effectively manage and organize a large amount of collected data, enabling us to analyze and extract valuable insights from the data.

\section{Related Datasets}
\label{dataset-comparision}

% We compare our Microlens dataset with several well-known datasets, as shown in Table~\ref{dataset-comparison}. In terms of modal richness, we provide raw text, images, audio, and videos, which is a feature that previous datasets lacked. Additionally, our dataset boasts a massive scale of interactions, reaching the billion-level mark, making it one of the largest datasets in the field of recommender systems. We are still expanding our Microlens dataset, aiming to facilitate research in various areas such as text recommendation, image recommendation, music recommendation, video recommendation, multi-modal recommendation, as well as the development of large-scale language models for recommender systems.

% In conclusion, as a novel dataset, Microlens stands out in terms of modal richness and interaction scale when compared to existing ones. It presents a unique opportunity for researchers to delve into various modalities and tackle the challenges in recommendation tasks using a vast amount of real-world user interactions.

\begin{table*}[htbp]
    \caption{Dataset comparison. ``p-Image'' refers to pre-extracted visual features from pre-trained visual encoders (such as ResNet), while ``r-Image''  refers to images with raw image pixels. ``Audio and Video'' means the original full-length audio and video content.
    % Note that (1) MovieLens has multiple versions, and we only show statistics for the version that includes movie URLs, i.e., ML25M; (2) apart from Tenrec and Alibaba CTR, there are many other popular recommendation datasets that do not have multimodal features, and they have not been reported here. 
    % \textcolor{blue}{I suggest you change darker colored symbols. }
    }
    \label{dataset-comparison}
    \centering
    % \vskip 0.12in
    \scalebox{0.72}{
    \begin{tabular}{cccccccccccc}
        \toprule
        \multirow{2}{*}{Dataset} & \multicolumn{5}{c}{Modality} & \multicolumn{3}{c}{Scale} & \multirow{2}{*}{Domain} & \multirow{2}{*}{Language} \\
        \cmidrule(lr){2-6}\cmidrule(lr){7-9}
        ~ & Text & p-Image & r-Image & Audio & Video &\#user &\#item &\#inter. & ~ & ~ & ~ \\
        \midrule
          Tenrec    &\textcolor{red}{\faTimes} &\textcolor{red}{\faTimes} &\textcolor{red}{\faTimes} &\textcolor{red}{\faTimes} &\textcolor{red}{\faTimes} &6.41M   &4.11M  &190.48M &News \& Videos  &\textcolor{red}{\faTimes} \\
        UserBehavior &\textcolor{red}{\faTimes} &\textcolor{red}{\faTimes} &\textcolor{red}{\faTimes} &\textcolor{red}{\faTimes} &\textcolor{red}{\faTimes} &988K  &4.16M   &100.15M  &E-commerce  &\textcolor{red}{\faTimes} \\ 
        Alibaba CTR      &\textcolor{red}{\faTimes} &\textcolor{red}{\faTimes} &\textcolor{red}{\faTimes} &\textcolor{red}{\faTimes} &\textcolor{red}{\faTimes} &7.96M  &66K   &15M  &E-commerce  &\textcolor{red}{\faTimes} \\ 
        Amazon       &\textcolor{green}{\faCheck} &-- &\textcolor{green}{\faCheck} &\textcolor{red}{\faTimes} &\textcolor{red}{\faTimes} &     20.98M   &   9.35M     &    82.83M   &E-commerce      &en \\
        POG          &\textcolor{green}{\faCheck} &-- &\textcolor{green}{\faCheck} &\textcolor{red}{\faTimes} &\textcolor{red}{\faTimes} &     3.57M   &   1.01M      &    0.28B    & E-commerce       & zh  \\
        MIND         &\textcolor{green}{\faCheck} &\textcolor{red}{\faTimes} &\textcolor{red}{\faTimes} &\textcolor{red}{\faTimes} &\textcolor{red}{\faTimes} &    1.00M    &   161K     &     24.16M   &News         &en \\
        H\&M         &\textcolor{green}{\faCheck} &-- &\textcolor{green}{\faCheck} &\textcolor{red}{\faTimes} &\textcolor{red}{\faTimes} &     1.37M   &   106K     &     31.79M   &E-commerce      &en \\
        BeerAdvocate &\textcolor{green}{\faCheck} &\textcolor{red}{\faTimes} &\textcolor{red}{\faTimes} &\textcolor{red}{\faTimes} &\textcolor{red}{\faTimes} &33K  &66K  &1.59M    &E-commerce      &en \\
        RateBeer     &\textcolor{green}{\faCheck} &\textcolor{red}{\faTimes} &\textcolor{red}{\faTimes} &\textcolor{red}{\faTimes} &\textcolor{red}{\faTimes} &40K  &110K   &2.92M    &E-commerce      &en \\
        Google Local &\textcolor{green}{\faCheck} &\textcolor{red}{\faTimes} &\textcolor{red}{\faTimes} &\textcolor{red}{\faTimes} &\textcolor{red}{\faTimes} &113.64M &4.96M   &666.32M  &E-commerce     &en \\
        Flickr       &\textcolor{red}{\faTimes} &\textcolor{green}{\faCheck} &\textcolor{red}{\faTimes} &\textcolor{red}{\faTimes} &\textcolor{red}{\faTimes} & 8K & 105K   &5.90M  &Social Media     &en \\
        Pinterest    &\textcolor{red}{\faTimes} &-- &\textcolor{green}{\faCheck} &\textcolor{red}{\faTimes} &\textcolor{red}{\faTimes} &    46K    &   880K     &   2.56M   &Social Media &\textcolor{red}{\faTimes} \\
        WikiMedia    &\textcolor{red}{\faTimes} &-- &\textcolor{green}{\faCheck} &\textcolor{red}{\faTimes} &\textcolor{red}{\faTimes} & 1K  & 10K & 1.77M &Social Media  &\textcolor{red}{\faTimes} \\
        Yelp         &\textcolor{red}{\faTimes} &-- &\textcolor{green}{\faCheck} &\textcolor{red}{\faTimes} &\textcolor{red}{\faTimes} &150K &200K   &6.99M   &E-commerce     &\textcolor{red}{\faTimes} \\
        GEST         &\textcolor{green}{\faCheck} &-- &\textcolor{green}{\faCheck} &\textcolor{red}{\faTimes} &\textcolor{red}{\faTimes} &1.01M   &4.43M   &1.77M  &E-commerce     &en \\
        Behance      &\textcolor{red}{\faTimes} &\textcolor{green}{\faCheck} &\textcolor{red}{\faTimes} &\textcolor{red}{\faTimes} &\textcolor{red}{\faTimes} &63K  &179K   &1.00M  &Social Media &\textcolor{red}{\faTimes} \\
        KuaiRand     &\textcolor{red}{\faTimes} &\textcolor{red}{\faTimes} &\textcolor{red}{\faTimes} &\textcolor{red}{\faTimes} &\textcolor{red}{\faTimes} &27K  &32.03M  &322.28M &Micro-video  &\textcolor{red}{\faTimes} \\
        KuaiRec      &\textcolor{red}{\faTimes} &\textcolor{green}{\faCheck} &\textcolor{red}{\faTimes} &\textcolor{red}{\faTimes} &\textcolor{red}{\faTimes} &7K   &11K  &12.53M  &Micro-video  &\textcolor{red}{\faTimes} \\
        ML25M    &\textcolor{green}{\faCheck} &-- &\textcolor{green}{\faCheck} &\textcolor{red}{\faTimes} &\textcolor{red}{\faTimes} &162K   &62K  &25.00M  &Movie-only  &en \\
        Reasoner     &\textcolor{green}{\faCheck} &-- &\textcolor{green}{\faCheck} &\textcolor{red}{\faTimes} &\textcolor{red}{\faTimes} & 3K  & 5K & 58K &Micro-video  &en \\
        \textbf{MicroLens}&\textcolor{green}{\faCheck} &-- &\textcolor{green}{\faCheck} &\textcolor{green}{\faCheck} &\textcolor{green}{\faCheck} &  30M & 1M & 1B &Micro-video  &zh/en \\
        \bottomrule
    \end{tabular}
    }
\end{table*}

\section{Hyper-parameter Settings for Baselines}
\label{baseline-parameter}
We report some essential hyperparameters of Baselines in Table~\ref{hyper-videorec}.
% The "Finetuning Layers" refers to finetuning the top blocks out four blocks in total in the video encoder network. 
% with an index greater than or equal to the given layer number. 
% For example, a value of 270 indicates that the bottom 270 out of the total 300 layers have been frozen. Non-linear and layer normalization are all regarded as separate layers. 
The  "finetuning rate" denotes the learning rate applied to the video encoder during the finetuning process.

% \begin{table*}[htbp]
%     \caption{Hyper-parameters settings for Baselines.}
%     \centering
%     \label{hyper-videorec}
%     % \vskip 0.12in
%     \scalebox{0.80}{
%     \begin{tabular}{ccccccccc}
%         \toprule
%         Model & \makecell[c]{Learning\\Rate} & \makecell[c]{Embedding\\Size} & \makecell[c]{Batch\\Size} & \makecell[c]{Dropout\\Rate} & \makecell[c]{Weight\\Decay} & Block Number & \makecell[c]{Finetuning\\Layers} & \makecell[c]{Finetuning\\Rate} \\
%         \midrule
%         NexItNet$_{\rm V}$ &1e-4  & 512 & 120 & 0.1 & 0.1 &2 (CNN Block) & 270 & 1e-4 \\
%         GRU4Rec$_{\rm V}$ & 1e-4 & 512 & 120 & 0.1 & 0.1 &1 (GRU Block) & 270 & 1e-4 \\
%         SASRec$_{\rm V}$ & 1e-4 & 512 & 120 & 0.1 & 0.1 & 2 (Transformer Block) & 270 & 1e-4 \\
%         \bottomrule
%     \end{tabular}
%     }
% \end{table*}

\begin{table*}[htbp]
    \caption{Hyper-parameters settings for baselines.}
    \centering
    \label{hyper-videorec}
    % \vskip 0.12in
    \scalebox{0.69}{
    \begin{tabular}{llcccccccc}
        \toprule
        Class & Model & \makecell[c]{Learning\\Rate} & \makecell[c]{Embedding\\Size} & \makecell[c]{Batch\\Size} & \makecell[c]{Dropout\\Rate} & \makecell[c]{Weight\\Decay} & Block Number & \makecell[c]{Finetune\\Top Blocks} & \makecell[c]{Finetuning\\Rate} \\
        \midrule
        \multirow{7}{*}{IDRec} & DSSM &1e-5  & 4096 & 64 & 0 & 0.1 &- & - & - \\
        & LightGCN &1e-3  & 1024 & 1024 & 0 & 0 &- & - & - \\
        & NFM &5e-5  & 1024 & 64 & 0 & 0.01 &- & - & - \\
        & DeepFM &1e-4  & 512 & 64 & 0 & 0.1 &- & - & - \\
        & NexItNet &1e-3  & 2048 & 64 & 0.1 & 0.1 &2 (CNN Block) & - & - \\
        & GRU4Rec & 1e-4 & 2048 & 512 & 0.1 & 0.1 &1 (GRU Block) & -& - \\
        & SASRec & 1e-5 & 2048 & 512 & 0.1 & 0.1 & 2 (Transformer Block) & - & - \\
        \midrule
        \multirow{8}{*}{VIDRec}   & Youtube$_{\rm ID}$ & 1e-4 & 4096 & 512 & 0.1 & 0.1 & - & - & - \\
        & Youtube$_{\rm ID+V}$ & 1e-4 & 4096 & 512 & 0.1 & 0.1 & - & - & - \\
        & MMGCN$_{\rm ID}$  & 1e-4 & 4096 & 64 & 0.1 & 0.0 & - & - & - \\
        & MMGCN$_{\rm ID+V}$ & 1e-4 & 4096 & 64 & 0.1 & 0.0 & - & - & - \\
        & GRCN$_{\rm ID}$& 1e-4 & 4096 & 64 & 0.1 & 0.0 & - & - & - \\
        & GRCN$_{\rm ID+V}$ & 1e-4 & 4096 & 64 & 0.1 & 0.0 & - & - & - \\
        & DSSM$_{\rm ID+V}$ & 1e-3 & 4096 & 1024 & 0 & 0.1 & - & - & - \\
        & SASRec$_{\rm ID+V}$ & 1e-5 & 2048 & 64 & 0.1 & 0.1 & - & - & - \\
        \midrule
        \multirow{3}{*}{VideoRec} & NexItNet$_{\rm V}$ &1e-4  & 512 & 120 & 0.1 & 0.1 &2 (CNN Block) & 1 & 1e-4 \\
        & GRU4Rec$_{\rm V}$ & 1e-4 & 512 & 120 & 0.1 & 0.1 &1 (GRU Block) & 1 & 1e-4 \\
        & SASRec$_{\rm V}$ & 1e-4 & 512 & 120 & 0.1 & 0.1 & 2 (Transformer Block) & 1 & 1e-4 \\

        \bottomrule
    \end{tabular}
    }
\end{table*}

\section{Video Model Details in Video Understanding and Recommendation}
\label{video-model-description}

\begin{table*}[htbp]
    \caption{Performance of VideoRec with 15 video encoders.
  "Pretrain Settings" are the adopted frame length and sample rate from the pre-trained checkpoint.  ACC@5 is the accuracy in the video classification task.
  % "VIT" is an abbreviation for Vision Transformer.
  }
    \label{video-models}
    \centering
    \scalebox{0.65}{
    \begin{tabular}{lcccccccccc}
        \toprule
        Model & Architecture & Depth & \makecell[c]{Pretrain\\Settings} & ACC@5 & \makecell[c]{HR@10\\(frozen)} & \makecell[c]{NDCG@10\\(frozen)} & \makecell[c]{HR@10\\(topT)} & \makecell[c]{NDCG@10\\(topT)} & \makecell[c]{HR@10\\(FT)} & \makecell[c]{NDCG@10\\(FT)} \\
        \midrule
        R3D-r18~\cite{tran2018closer} & ResNet & R18 & 16x4 & 75.45 & 4.58 & 2.56 & 8.50 & 4.48 & 7.50 & 3.48 \\
        X3D-xs~\cite{feichtenhofer2020x3d} & Xception & XS & 4x12 & 88.63 & 0.62 & 0.33 & 7.04 & 3.57 & 6.04 & 2.57 \\
        C2D-r50~\cite{wang2018non} & ResNet & R50 & 8x8 & 89.68 & 4.11 & 2.27 & 9.22 & 4.88 & 8.22 & 3.88 \\
        I3D-r50~\cite{carreira2017quo} & ResNet & R50 & 8x8 & 90.70 & 4.19 & 2.36 & 9.25 & 5.01 & 8.25 & 4.01 \\
        X3D-s~\cite{feichtenhofer2020x3d} & Xception & S & 13x6 & 91.27 & 0.47 & 0.24 & 6.43 & 3.25 & 5.43 & 2.25 \\
        Slow-r50~\cite{fan2021pytorchvideo} & ResNet & R50 & 8x8 & 91.63 & 4.42 & 2.42 & 9.32 & 4.99 & 8.33 & 3.99 \\
        X3D-m~\cite{feichtenhofer2020x3d} & Xception & M & 16x5 & 92.72 & 0.38 & 0.20 & 6.11 & 3.13 & 5.11 & 2.13 \\
        R3D-r50~\cite{tran2018closer} & ResNet & R50 & 16x4 & 92.23 & 0.28 & 0.14 & 8.33 & 4.34 & 7.33 & 3.34 \\
        SlowFast-r50~\cite{feichtenhofer2019slowfast} & ResNet & R50 & 8x8 & 92.69 & 4.14 & 2.35 & 9.48 & 5.15 & 8.48 & 4.15 \\
        CSN-r101~\cite{tran2019video} & ResNet & R101 & 32x2 & 92.90 & 4.48 & 2.52 & 8.74 & 4.71 & 7.74 & 3.71 \\
        X3D-l~\cite{feichtenhofer2020x3d} & Xception & L & 16x5 & 93.31 & 0.64 & 0.34 & 6.37 & 3.32 & 5.37 & 2.32 \\
        SlowFast-r101~\cite{feichtenhofer2019slowfast} & ResNet & R101 & 16x8 & 93.61 & 4.25 & 2.36 & \textbf{9.76} & \textbf{5.3} & \textbf{8.76} & \textbf{4.31} \\
        MViT-B-16x4~\cite{fan2021multiscale} & VIT & B & 16x4 & 93.85 & 2.30 & 1.33 & 8.96 & 4.79 & 7.96 & 3.79 \\
        MViT-B-32x3~\cite{fan2021multiscale} & VIT & B & 32x3 & 94.69 & 1.95 & 1.11 & 9.57 & 5.11 & 8.57 & 4.11 \\
        VideoMAE~\cite{tong2022videomae} & Transformer & VIT-B & 16x4 & \textbf{95.10} & \textbf{4.96} & \textbf{2.76} & 8.91 & 4.77 & 7.91 & 3.77 \\
        \bottomrule
    \end{tabular}
    }
\end{table*}

\section{Details of the Applied Image Encoders}

We showed details of three classical image encoders in Table~\ref{MEversion}.
% Note that the upstream checkpoints we used are indicated in the "URL" column.

\begin{table}[htbp]
   \caption{Network architecture, parameter size, and download URL of the vision encoders for image baselines. L: number of Transformer blocks, H: number of multi-head attention, C: channel number of the hidden layers in the first stage, B: number of layers in each block.
   }
   \label{MEversion}
   \begin{center}
   \scalebox{0.8}{
               \begin{tabular}{  c  c  c  c  c }%
               \toprule
               Image encoder & Architecture & \#Param.  & URL \\
               \midrule
               ResNet18 & C = 64, B=\{2, 2, 2, 2\} & 12M  & https://download.pytorch.org/models/resnet18-5c106cde.pth\\
               Swin-T   & C = 96, B=\{2, 2, 6, 2\} & 28M  & https://huggingface.co/microsoft/swin-tiny-patch4-window7-224\\
               MAE$_{base}$    & L=12, H=768 & 86M & https://huggingface.co/facebook/vit-mae-base\\
               \bottomrule
               \end{tabular}}
   \end{center}
\end{table}

\section{Warm-up Recommendation on MicroLens-100K}
% The results from Table~\ref{warmup-table} showcase that VideoRec significantly outperforms IDRec in both regular and  warm-up scenarios. Even in the warm-up (200) setting, the difference between the two methods remains pronounced. It is a surprising finding since IDRec is known for its ability to model popular items, and the experiments conducted in \cite{yuan2023go} suggest that IDRec should have a slight advantage over modality-based methods in warmer settings. However, it is important to note that the modal content used in \cite{yuan2023go} are covers or titles, instead of video media. The comparison verifies that compared to images or text, video content indeed provides richer item representation information and thus amplifies the advantages of modality-based methods, again highlighting the merits of the proposed MicroLens. 

\begin{table*}[htbp]
    \caption{Comparison of VideoRec and IDRec in regular and warm settings using SASRec as the backbone. 
     ``Warm-20'' denotes that items with less than 20 interactions were removed from the original MicroLens-100K.
    }
    \label{warmup-table}
    \centering
    % \vskip 0.12in
    \scalebox{0.90}{
    \begin{tabular}{ccccccccc}
        \toprule
        \multirow{3}{*}{Model} & \multicolumn{2}{c}{Regular} & \multicolumn{2}{c}{Warm-20} & \multicolumn{2}{c}{Warm-50} & \multicolumn{2}{c}{Warm-200}\\
        \cmidrule(lr){2-3} \cmidrule(lr){4-5} \cmidrule(lr){6-7} \cmidrule(lr){8-9}
        % & \multicolumn{2}{c}{MIND} & \multicolumn{2}{c}{HM} & \multicolumn{2}{c}{BILI} & \multicolumn{2}{c}{MIND} & \multicolumn{2}{c}{HM} & \multicolumn{2}{c}{BILI} \\
        % \cmidrule(lr){2-3} \cmidrule(lr){4-5} \cmidrule(lr){6-7}\cmidrule(lr){8-9} \cmidrule(lr){10-11} \cmidrule(lr){12-13}
        % & HR@10 & NDCG@10 & HR@20 & NDCG@20 & HR@10 & NDCG@10 & HR@20 & NDCG@20 \\
        & H@10 & N@10 & H@10 & N@10 & H@10 & N@10 & H@10 & N@10 \\
        \midrule
        IDRec & 0.0909 & 0.0517 & 0.1068 & 0.0615 & 0.6546 & 0.4103 & 0.7537 & 0.4412  \\
        SlowFast-r101      & 0.0976 & 0.0531 & 0.1130 & 0.0606 & 0.7458 & 0.4463 & 0.8482 & 0.4743 \\
        MViT-B-32x3  & 0.0957 & 0.0511 & 0.1178 & 0.0639 & 0.7464 & 0.4530 & 0.9194 & 0.4901 \\
        SlowFast-r50     & 0.0948 & 0.0515 & 0.1169 & 0.0642 & 0.7580 & 0.4614 & 0.8141 & 0.4870 \\
        \bottomrule
    \end{tabular}
    }
\end{table*}

\section{Baseline Evaluation and Warm-up Recommendation on MicroLens-1M}
We reported the results of baseline evaluation and warm-up recommendation on MicroLens-1M in Table~\ref{baseline-large-table} and Table~\ref{warmup-large-table}, respectively. 
Please note that due to excessive GPU memory consumption, some baselines could not be trained on MicroLens-1M, and we do not report their  results. In general, we observed that the trends on  MicroLens-1M (in terms of both baseline evaluation and warm-up recommendation) are consistent with that observed on  MicroLens-100K.
% , which demonstrates the solidity of our findings and analysis.

\begin{table*}[htbp]
    \caption{Benchmark results on MicroLens-1M. 
    % VideoMAE and SlowFast are used as video encoder for VIDRec and VideoRec, respectively.
    }
    \label{baseline-large-table}
    \centering
    % \vskip 0.12in
    \scalebox{0.90}{
    \begin{tabular}{llcccc}
        \toprule
        % \multirow{3}{*}{Class} & \multirow{3}{*}{Model} & \multicolumn{4}{c}{MicroLens-100K} & \multicolumn{4}{c}{MicroLens-1000K} \\
        Class & Model & HR@10 & NDCG@10 & HR@20 & NDCG@20  \\
        % & \multicolumn{2}{c}{MIND} & \multicolumn{2}{c}{HM} & \multicolumn{2}{c}{BILI} & \multicolumn{2}{c}{MIND} & \multicolumn{2}{c}{HM} & \multicolumn{2}{c}{BILI} \\
        % \cmidrule(lr){2-3} \cmidrule(lr){4-5} \cmidrule(lr){6-7}\cmidrule(lr){8-9} \cmidrule(lr){10-11} \cmidrule(lr){12-13}
        % & HR@10 & NDCG@10 & HR@20 & NDCG@20 & HR@10 & NDCG@10 & HR@20 & NDCG@20 \\
        % & & H@10 & N@10 & H@20 & N@20 & H@10 & N@10 & H@20 & N@20 \\
        \midrule
        \multirow{2}{*}{IDRec (CF)} 
        %& FM      &   0.0215     & 0.0107  &  0.0348  &  0.0141       \\
        & DSSM      & 0.0133 & 0.0065 & 0.0225 & 0.0087 \\
        & LightGCN  & 0.0150& 0.0072 & 0.0253 & 0.0098\\
\midrule
\multirow{3}{*}{IDRec (SR)} 
        & NexItNet & 0.0389& 0.0209 & 0.0584 & 0.0258 \\
        & GRU4Rec  & 0.0444 & 0.0234 & 0.0683 & 0.0294 \\
        & SASRec   & 0.0476 & 0.0255 & 0.0710 & 0.0314 \\
\midrule
\multirow{2}{*}{\makecell[l]{VIDRec\\(Frozen Encoder)}}
        & YouTube$_{\rm ID}$ &0.0256 & 0.0129  & 0.0578   &   0.0246    \\
        & YouTube$_{\rm ID+V}$ &0.0180    &0.0089   & 0.0303   &  0.0119     \\
        % & MMGCN$_{\rm ID}$ &        &        &        &        \\
        % & MMGCN$_{\rm ID+V}$ &        &        &        &        \\
        % & GRCN$_{\rm ID}$ &        &        &        &        \\
        % & GRCN$_{\rm ID+V}$ &        &        &        &        \\
        \midrule
        \multirow{3}{*}{\makecell[l]{VideoRec\\(E2E Learning)}}
        & NexItNet$_{\rm V}$  & 0.0521 & 0.0272 &0.0792 & 0.0340 \\
        & GRU4Rec$_{\rm V}$  & 0.0510 & 0.0264 &0.0782 & 0.0332 \\
        & SASRec$_{\rm V}$  & 0.0582 & 0.0309 & 0.0871 & 0.0382 \\
        \bottomrule
    \end{tabular}
    }
\end{table*}

% \section{Warm-up Recommendation on MicroLens-1M}
\begin{table*}[htbp]
    \caption{Comparison of VideoRec and IDRec in regular and warm-start settings using SASRec as the user backbone. Warm-20 denotes that items with less than 20 interactions were removed from the original MicroLens-1M.}
    \label{warmup-large-table}
    \centering
    % \vskip 0.12in
    \scalebox{0.90}{
    \begin{tabular}{ccccccccc}
        \toprule
        \multirow{3}{*}{Model} & \multicolumn{2}{c}{Regular} & \multicolumn{2}{c}{Warm-20} & \multicolumn{2}{c}{Warm-50} & \multicolumn{2}{c}{Warm-200}\\
        \cmidrule(lr){2-3} \cmidrule(lr){4-5} \cmidrule(lr){6-7} \cmidrule(lr){8-9}
        % & \multicolumn{2}{c}{MIND} & \multicolumn{2}{c}{HM} & \multicolumn{2}{c}{BILI} & \multicolumn{2}{c}{MIND} & \multicolumn{2}{c}{HM} & \multicolumn{2}{c}{BILI} \\
        % \cmidrule(lr){2-3} \cmidrule(lr){4-5} \cmidrule(lr){6-7}\cmidrule(lr){8-9} \cmidrule(lr){10-11} \cmidrule(lr){12-13}
        % & HR@10 & NDCG@10 & HR@20 & NDCG@20 & HR@10 & NDCG@10 & HR@20 & NDCG@20 \\
        & H@10 & N@10 & H@10 & N@10 & H@10 & N@10 & H@10 & N@10 \\
        \midrule
        IDRec     & 0.0476 & 0.0255 & 0.0508 & 0.0272 & 0.0562 & 0.0306 & 0.5533 & 0.3105 \\
        SlowFast-r101 & 0.0554 & 0.0291 & 0.0574 & 0.0303 & 0.0603 & 0.0318 & 0.5766 & 0.3107 \\
        MViT-B-32x3  & 0.0569 & 0.0300 & 0.0562 & 0.0294 & 0.0608 & 0.0318 & 0.6046 & 0.3399 \\
        SlowFast-r50    & 0.0582 & 0.0309 & 0.0603 & 0.0319 & 0.0638 & 0.0339 & 0.6217 & 0.3556 \\
        \bottomrule
    \end{tabular}
    }
\end{table*}

% \section{Finetune settings for video models.}
% \label{video-settings}

% \begin{table*}[htbp]
%     \caption{Finetune settings for video models.}
%     \label{video-parameter-table}
%     \centering
%     % \vskip 0.12in
%     \scalebox{0.90}{
%     \begin{tabular}{ccccccccc}
%         \toprule
%         % \multirow{3}{*}{Model} & \multicolumn{2}{c}{Regular} & \multicolumn{2}{c}{Warm-up (20)} & \multicolumn{2}{c}{Warm-up (50)} & \multicolumn{2}{c}{Warm-up (200)}\\
%         % \cmidrule(lr){2-3} \cmidrule(lr){4-5} \cmidrule(lr){6-7} \cmidrule(lr){8-9}
%         % % & \multicolumn{2}{c}{MIND} & \multicolumn{2}{c}{HM} & \multicolumn{2}{c}{BILI} & \multicolumn{2}{c}{MIND} & \multicolumn{2}{c}{HM} & \multicolumn{2}{c}{BILI} \\
%         % % \cmidrule(lr){2-3} \cmidrule(lr){4-5} \cmidrule(lr){6-7}\cmidrule(lr){8-9} \cmidrule(lr){10-11} \cmidrule(lr){12-13}
%         % % & HR@10 & NDCG@10 & HR@20 & NDCG@20 & HR@10 & NDCG@10 & HR@20 & NDCG@20 \\
%         % & H@10 & N@10 & H@10 & N@10 & H@10 & N@10 & H@10 & N@10 \\
%         % \midrule
%         % ID      & 0.0476 & 0.0255 & 0.0508 & 0.0272 & 0.0562 & 0.0306 & 0.5533 & 0.3105 \\
%         % SlowFast-r101 & 0.0554 & 0.0291 & 0.0574 & 0.0303 & 0.0603 & 0.0318 & 0.5766 & 0.3107 \\
%         % MViT-B-32x3  & 0.0569 & 0.0300 & 0.0562 & 0.0294 & 0.0608 & 0.0318 & 0.6046 & 0.3399 \\
%         % SlowFast-r50    & 0.0582 & 0.0309 & 0.0603 & 0.0319 & 0.0638 & 0.0339 & 0.6217 & 0.3556 \\
%         \bottomrule
%     \end{tabular}
%     }
% \end{table*}

\section{Recommendation in Cold-start Scenarios}
\label{cold-start-section}
% We analyze the performance of SASRec (IDRec) and SASRec$_{V}$ with three video models (VideoRec) under cold-start scenarios. Specifically, items are grouped based on their frequency of occurrence (i.e., item popularity), with varying levels representing different degrees of cold-start situations.
% It should be noted that the 0-popularity signifies that all items involved in the test set are novel and have not previously been encountered by the models.

% The results presented in Figure~\ref{cold-start} demonstrate a significant disparity between VideoRec and IDRec in cold-start scenarios. IDRec relies on interactions between users and items to learn their representations, while VideoRec can model items based on the content of videos themselves. This enables VideoRec to provide more pronounced recommendation results even in situations with minimal interactions compared to IDRec. Moreover, as the cold-start problem becomes more severe, the advantage of VideoRec over IDRec becomes increasingly amplified. For instance, as the popularity becomes smaller, the relative improvement of VideoRec to IDRec larger, especially when the popularity varies from 2 to 1.
% For instance, when new items emerge, IDRec is unable to make recommendations since they are unknown to the model, while VideoRec can leverage knowledge from other video content to understand these new items. 
% Therefore, the  content in MicroLens would also contribute to advancing research on the cold-start problem.

\begin{figure*}[h]
    \centering
    \includegraphics[width=0.5\textwidth]{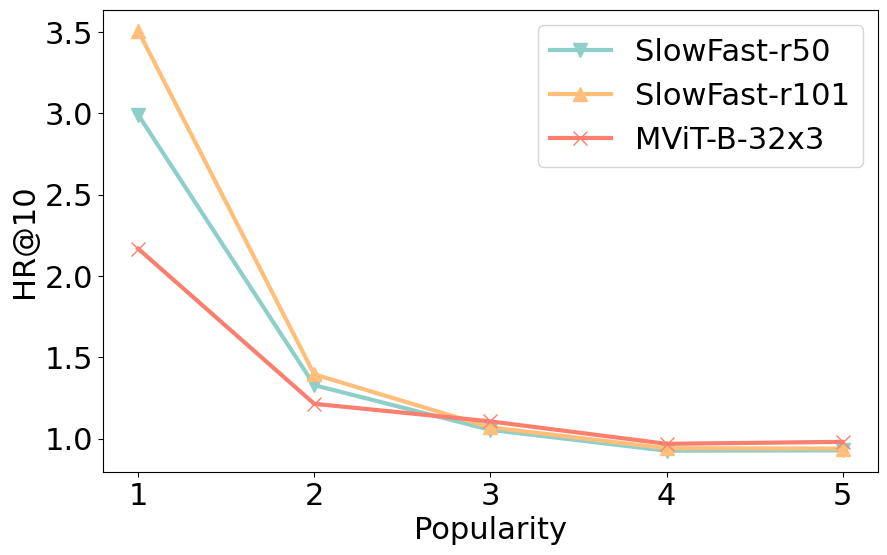}
    \caption{Results in different cold-start scenarios, with the y-axis representing the relative improvement of HR@10, calculated as the ratio of VideoRec to IDRec.  The x-axis represents item groups divided by popularity level, the larger number indicates that items in the group are more popular.}
    \label{cold-start}
\end{figure*}

\section{Recommendation with Side Features}
\label{side-feature-section}

\begin{table*}[htbp]
    \caption{Recommendation results with side features on MicroLens-100K. 
    }
    \label{side-feature-table}
    \centering
    % \vskip 0.12in
    \scalebox{1.0}{
    \begin{tabular}{lcccc}
        \toprule
        % \multirow{3}{*}{Class} & \multirow{3}{*}{Model} & \multicolumn{4}{c}{MicroLens-100K} & \multicolumn{4}{c}{MicroLens-1000K} \\
         Model & HR@10 & NDCG@10 & HR@20 & NDCG@20  \\
        \midrule
        SASRec$_{\rm ID}$      & 0.0909 & 0.0517 & 0.1278 & 0.0610 \\
        SASRec$_{\rm ID+Pop}$  & 0.0709 & 0.0396 & 0.1037 & 0.0479\\
        SASRec$_{\rm ID+Tag}$  & 0.0908 & 0.0499 & 0.1320 & 0.0603 \\
        SASRec$_{\rm ID+Pop+Tag}$  & 0.0778 & 0.0423 &0.1138 & 0.0513 \\
        \bottomrule
    \end{tabular}
    }
\end{table*}

In this section, we investigate the impact of other features on recommendation performance using MicroLens-100K dataset. We introduce two types of side features: item popularity level (Pop) and tag categories (Tag). For popularity features, we divide the item popularity into 10 uniform bins. The first bin represents the top 10\% of popular items, while the last bin represents the bottom 10\%. We assign a Pop ID to each item according to its popularity level. Regarding the tag features, we also handle them as categorical features with a category of $15,580$. 
% We then concatenate the tag ID embedding, popularity ID embedding, and corresponding item ID embedding as the final item embedding.
% we first assign corresponding IDs to all tags in MicroLens-100K and then assign 0 to multiple Tag IDs to items based on the tag categories associated with each item.
% To utilize the side features, we create learnable embeddings for Pop ID and Tag ID, with a vocabulary size of 10 and 15580 respectively.
%To learn the side features, we create learnable matrices for Pop ID and Tag ID, with dimensions of $ed$x$10$ and $ed$x$15580$, respectively. Here, $ed$ represents the embedding dimension of Item ID, Pop ID, and Tag ID. 
% The number $10$ represents the level number of item popularity, and $15,580$ is the tag count in MicroLens-100K.

We conducted experiments on SASRec$_{\rm ID}$ (ID) with different feature combinations: SASRec$_{\rm ID}$, SASRec$_{\rm ID+Pop}$, SASRec$_{\rm ID+Tag}$, and SASRec$_{\rm ID+Pop+Tag}$. The "+" symbol denotes feature combination achieved by summing and averaging them. We report the results in Table~\ref{side-feature-table}.

We found that incorporating item popularity level and tag categories as side features did not clearly improve the algorithm's performance. One possible reason is that in typical recommendation scenarios, item ID embeddings have already been extensively trained, implicitly learning latent factors including similarity and popularity. For instance, we observed that many videos recommended in the top-10 recommendation list share similar categories and have relatively high popularity, indicating that ID-based methods can already capture popularity and category information. In such scenarios, incorporating many unimportant features  may  have a negative impact on overall performance. It is worth noting that in the very cold-start setting, the item ID feature is very weak and adding other features is necessary for better performance.

\section{Comparison between Textual Features and Video Content}
\label{text-video-comparision-section}

\begin{table*}[htbp]
    \caption{Comparsion results of ID, textual features and video content on MicroLens-100K.  
    }
    \label{text-video-table}
    \centering
    % \vskip 0.12in
    \scalebox{1.0}{
    \begin{tabular}{lcccc}
        \toprule
        % \multirow{3}{*}{Class} & \multirow{3}{*}{Model} & \multicolumn{4}{c}{MicroLens-100K} & \multicolumn{4}{c}{MicroLens-1000K} \\
         Model & HR@10 & NDCG@10 & HR@20 & NDCG@20  \\
        \midrule
        SASRec$_{\rm ID}$      & 0.0909 & 0.0517 & 0.1278 & 0.0610 \\
        SASRec$_{\rm T}$  & 0.0916 & 0.0490 & 0.1343 & 0.0598 \\
        SASRec$_{\rm V}$  & 0.0953 & 0.0520 & 0.1374 & 0.0626 \\
        \bottomrule
    \end{tabular}
    }
\end{table*}

% Text features are widely used as important modal information in modality-based recommendation. However, for micro videos, the information contained in the video content is often richer than that of textual titles. To validate this claim, we employed SASRec as the recommender backbone and investigated the impact of ID embedding (ID), textual titles (T), and video content (V) on recommendation performance.

We used BERT\footnote{https://huggingface.co/prajjwal1/bert-small} as the text encoder and SlowFast16x8-r101 as the video encoder and perform end-to-end training as mentioned in section 3.1. We fixed the learning rate of recommender model as $1e-4$, and searched for the optimal learning rates for the text encoder and video encoder from $\{1e-3, 1e-4\}$. The comparison results are reported in Table~\ref{text-video-table}.
Our results demonstrate that using only text features yields similar performance to  the itemID feature. By analyzing the data, we have observed that some short videos have only a few words in their descriptions, which may contribute to the performance not being particularly competitive. On the other hand, the amount of information contained in the original  videos far exceeds that of the video titles. Therefore, we believe that in the future, utilizing more powerful video understanding techniques can lead to  better recommendation results.

% \textcolor{blue}{
% From the results, we found that recommendation with video and text features are better than that with item ID embeddings. Furthermore, we observe that video and text features are comparable in terms of performance, with video slightly outperforming text. However, the improvement in video performance is not significant, which could be attributed to the fact that the current state of video understanding techniques is still in an early stage compared to language-based models (LLM). We believe that as video understanding technology advances, video-based recommendation algorithms will play a more important role in short video recommendation.
% }

\section{Recommendation Scenario of Collected Platform}
\label{recommendation-scenario}

\begin{figure*}[h]
    % \captionsetup{labelfont={color=blue}}
    \centering
    \includegraphics[width=0.8\textwidth]{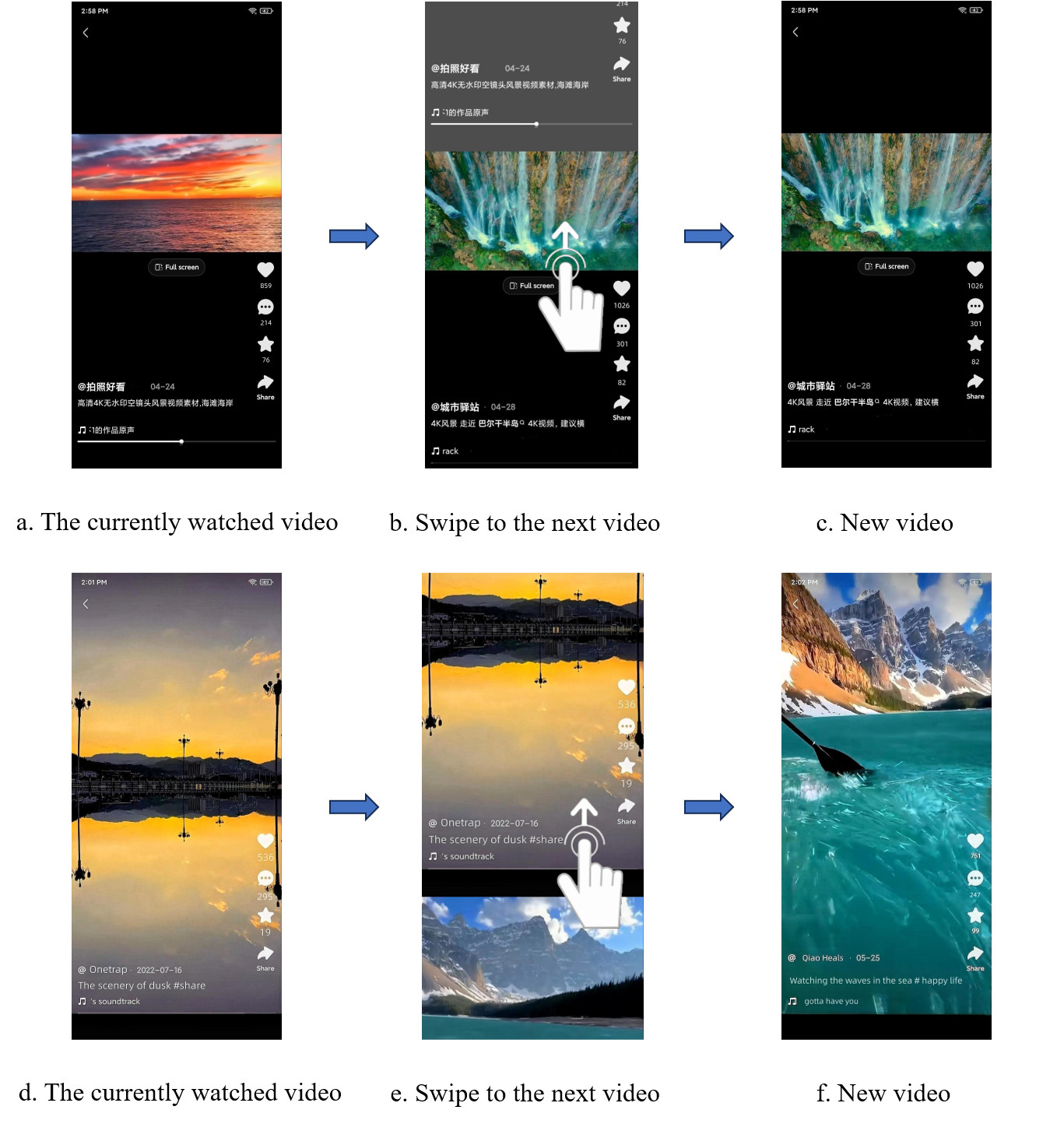}
    \caption{An illustration of the recommendation scenario in MicroLens. Videos a, b, and c are displayed in landscape format, while videos d, e, and f are displayed in portrait format. Please note that the format of the next video is random and can be either landscape or portrait. English translation is provided for all video titles.}
    \label{recommendation-scenario-figure}
\end{figure*}

Figure~\ref{recommendation-scenario-figure} illustrates the recommendation scenario of the micro video platform from which our MicroLens collected data. In this example, a user is recommended a video about trucks. After watching a short segment, the user swipes up to the next video. All these videos allow user engagement through buttons for liking, sharing, and commenting, which are visible on the right side of the videos. On this platform, there are multiple ways to define positive and negative examples. For instance, the duration of video views, presence of likes, comments, or shares can all be considered as different levels of user feedback.  However, among these behaviors, only comment behaviors are public without any access restrictions. Also,  note that the videos and comments are publicly accessible both on the mobile app and the web. In the mobile app, users navigate to the next video by swiping gestures, while on the web, users use mouse scrolling to move to the next video. The web scene is displayed in the same way as the mobile app scene.

In the micro-video application, users are typically presented with a continuous stream of videos. The recommendation process continues uninterrupted  through the user swiping up or mouse scrolling, ensuring a seamless flow of video recommendations.

\clearpage

\end{document}